\documentclass[fleqn,twoside]{article}

\usepackage{espcrc2}

\usepackage{graphicx}

\usepackage[figuresright]{rotating}

\newcommand{\AmS}{{\protect\the\textfont2

  A\kern-.1667em\lower.5ex\hbox{M}\kern-.125emS}}

\title{$\Delta\Delta$ Excitation in Proton-Proton Induced  $\pi^0\pi^0$
  Production }
 
\author{T.~Skorodko\address[PIT]{Physikalisches Institut der Universit\"at
  T\"ubingen, D-72076 T\"ubingen, Germany},
M.~Bashkanov\addressmark[PIT],
 D.~Bogoslawsky\address[JINR]{Joint Institute for Nuclear Research, Dubna,
  Russia},
H.~Cal\'en\address[SL]{The Svedberg Laboratory, Uppsala, Sweden},
H.~Clement\addressmark[PIT],
E.~Doroshkevich\addressmark[PIT],
L.~Demiroers\address[HU]{Hamburg University, Hamburg, Germany},
C.~Ekstr\"om\addressmark[SL],
K.~Fransson\addressmark[SL],
L.~Gustafsson\address[UU]{Uppsala University, Uppsala,Sweden},
B.~H\"oistad\addressmark[UU],
G.~Ivanov\addressmark[JINR],
M.~Jacewicz\addressmark[UU],
E.~Jiganov\addressmark[JINR],
T.~Johansson\addressmark[UU],
O.~Khakimova\addressmark[PIT],
S.~Keleta\addressmark[UU],
I.~Koch\addressmark[UU],
F.~Kren\addressmark[PIT],
S.~Kullander\addressmark[UU],
A.~Kup\'s\'c\addressmark[SL],
P.~Marciniewski\addressmark[SL],
R.~Meier\addressmark[PIT],
B.~Morosov\addressmark[JINR],
C.~Pauly\address[FJ]{Forschungszentrum J\"ulich, Germany},
H.~Petr{\'e}n.\addressmark[UU],
Y.~Petukhov\addressmark[JINR],
A.~Povtorejko\addressmark[JINR],
R.J.M.Y.~Ruber\addressmark[SL],
K.~Sch\"onning\addressmark[UU],
W.~Scobel\addressmark[HU],
B.~Shwartz\address[BINP]{Budker Institute of Nuclear Physics, Novosibirsk,
  Russia},
J.~Stepaniak\address[SINS]{Soltan Institute of Nuclear Studies, Warsaw and
  Lodz, Poland},
P.~Th\"orngren-Engblom\address[SU]{Stockholm University, Stockholm, Sweden}
V.~Tikhomirov\addressmark[JINR],
G.J.~Wagner\addressmark[PIT],
M.~Wolke\addressmark[UU],
A.~Yamamoto\address[HEARO]{High Energy Accelerator Research Organization,
  Tsukuba, Japan},
 J.~Zabierowski\addressmark[SINS],
and
J.~Zlomanczuk\addressmark[UU]}

\begin{document}

\begin{abstract}

Exclusive measurements of the $pp \to pp\pi^0\pi^0$ reaction have been
performed at CELSIUS/WASA at energies from threshold up to $T_p$ = 1.3
GeV. Total and differential cross sections have been obtained. Here we
concentrate on energies $T_p \ge$ 1 GeV, where the $\Delta\Delta$ excitation
becomes the leading process. No evidence is found for a significant ABC effect 
beyond that given by the conventional $t$-channel $\Delta\Delta$
excitation. This holds also for the double-pionic fusion to the quasibound
$^2$He. The data are compared to model predictions, which are based
on both pion and $\rho$ exchange. Total and differential cross sections 
are at variance with these predictions and call for a profound modification of
the $\rho$-exchange. A phenomenological modification allowing only a small
$\rho$ exchange contribution leads to a quantitative description
of the data. 
\vspace{1pc}

\end{abstract}


\maketitle

Two-pion production in nucleon-nucleon collisions connects $\pi\pi$ dynamics
with baryon and baryon-baryon degrees of freedom. In the special case that the
participating nucleons fuse to a bound nuclear system, there is the
puzzling ABC effect, which stands for a low-mass enhancement in the isoscalar
$\pi\pi$ invariant mass spectrum. Very recent experiments on this topic are
discussed in terms of 
a $\Delta\Delta$ mediated isoscalar resonance in the baryon-baryon system as
source for this peculiar ABC effect\cite{bash,MB,hcl}. 

By contrast the isovector $\pi\pi$ 
channel in double-pionic fusion behaves regularly, i.e. shows no ABC effect
and follows the expectations from conventional t-channel $\Delta\Delta$
calculations \cite{FK}. Also in the two-pion production to unbound nuclear
systems the ABC effect was thought to be absent. However, a very recent
inclusive measurement of the reaction $pp \to ppX$, where X stands for ejectiles
not detected in the experiment, reports evidence for an ABC effect also in
this case \cite{dymov}. 

A recent isospin decomposition \cite{tsi} of the total cross sections measured
in the reactions $pp \to pp\pi^+\pi^-$, $pp \to pp\pi^0\pi^0$, $pp \to
pn\pi^+\pi^0$ and $pp \to nn\pi^+\pi^+$ reveal these two-pion production
channels to be dominated by excitation and decay of resonances. In particular
the $N^*(1440)$ dominates at energies close to threshold and the
$\Delta\Delta$ system and possibly the $\Delta(1600)$ at higher incident
energies. The latter is supposed to contribute primarily to the $pp \to
nn\pi^+\pi^+$ and also to the $pp \to pn\pi^+\pi^0$ channel.

In view of the challenging interpretation \cite{MB} offered for the ABC effect
in isoscalar $\pi\pi$ channels in case of double-pionic fusion and the reported
evidence in the inclusive $pp \to ppX$ reaction it appears mandatory 
to study the isoscalar $\pi\pi$ production with exclusive and kinematically
complete measurements in the case, where the two
participating nucleons do not fuse into a final nuclear bound system. Among the
two possible choices, the  $pp\pi^+\pi^-$ or the  $pp\pi^0\pi^0$ channel, the
latter one is especially appealing, since it contains no $\pi\pi$ isovector
contributions, only isoscalar and isotensor parts with the isoscalar part being 
the by far dominating one \cite{tsi}

From previous work it is known that the
$pp \to pp\pi^+\pi^-$ and $pp \to pp\pi^0\pi^0$ reactions in the
near-threshold region are well understood as being dominated by excitation and
decay of the Roper resonance \cite{alv,WB,JP,skor}. At higher energies
theoretical calculations \cite{alv} predict the t-channel 
$\Delta\Delta$ excitation to play the dominant part. These calculations are
compared in Figs. 1 - 5 with the differential and total cross section data for
the $pp \to pp\pi^0\pi^0$ reaction obtained in this work.

Since there have been no exclusive measurements of the $pp \to pp\pi^0\pi^0$
channel in the energy  region of interest, we have carried out a systematic
program of exclusive two-pion production measurements in $pp$ collisions from
threshold up to $T_p$ = 1.36 GeV using the WASA detector  
\cite{barg} with the hydrogen pellet target system at the CELSIUS storage
ring of the The Svedberg Laboratory in Uppsala. The detector has nearly
full angular coverage 
for the detection of charged particles and photons. The forward detector
consists of a thin-walled window plastic scintillator hodoscope at the exit of
the scattering chamber, followed by straw tracker, plastic scintillator quirl
and range hodoscopes, whereas the central detector comprises an
electromagnetic calorimeter consisting of 1012 CsI (Na) crystals, and in its
inner part a plastic scintillator barrel surrounding a thin-walled
superconducting magnet containing a mini drift chamber for tracking. 

Neutral pions are reconstructed from photons detected and identified in
the central detector. Protons are detected in the forward detector and
identified by the $\Delta$E-E technique. Since the forward detector cone does
not cover the full kinematic angular range for protons at high incident
energies, the detection efficiency for protons at medium center-of-mass (cms)
angles is reduced. This introduces systematic uncertainties in particular in
the proton angular distribution. Since due to the identity of the two incident
particles, the angular distributions have to be symmetric about
90$^{\circ}$, the observed asymmetries about 90$^{\circ}$ (see Figs. 3 and 4)
may hence serve as a measure of such systematic errors in the data. 
We estimate the systematic uncertainties due to this deficiency of full phase
space coverage by using various model calculations in the Monte Carlo
simulations of the detector response and acceptance corrections. The estimated
systematic uncertainties are shown by dark-shaded histograms in Figs.~2 - 5. 

The absolute normalization of the data has been achieved by a simultaneous
measurement of elastic scattering and/or single pion production, for which the
cross sections are known. Since in particular the single-pion production cross
sections are not known better than to an accuracy of 20 $\%$, this uncertainty
transmits also to the cross sections deduced for the $pp \to pp\pi^0\pi^0$
reaction. 
For details of the data analysis see, {\it e.g.} Refs. \cite{FK,tsi}. 

Experimental results for the low-energy range $T_p <$ 1 GeV have been
published \cite{skor} in connection with the discussion of the
properties of the Roper resonance. In addition close-to-threshold results
from previous PROMICE/WASA measurements are given in Ref. \cite{jan}.

\begin{figure} [t]
\begin{center}
\includegraphics[width=0.45\textwidth]{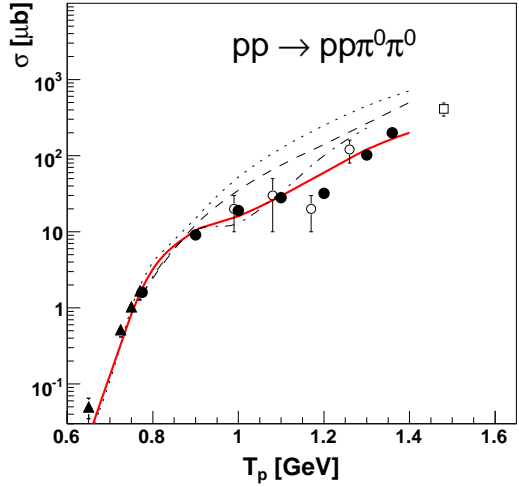}
\caption{ 
   Total cross sections for the $pp \to pp\pi^0\pi^0$ 
  reaction. Data are represented by open (bubble chamber data,
  Refs. \cite{shim,eis}) and filled symbols (WASA data,
  Refs. \cite{tsi,jan}). The dotted lines show the original calculation of 
   Ref. \cite{alv}. The dashed, dash-dotted and solid lines are calculations
   with modifications described in the text. 
}
\label{fig3}
\end{center}
\end{figure}

\begin{figure}
\begin{center}

\includegraphics[width=0.23\textwidth]{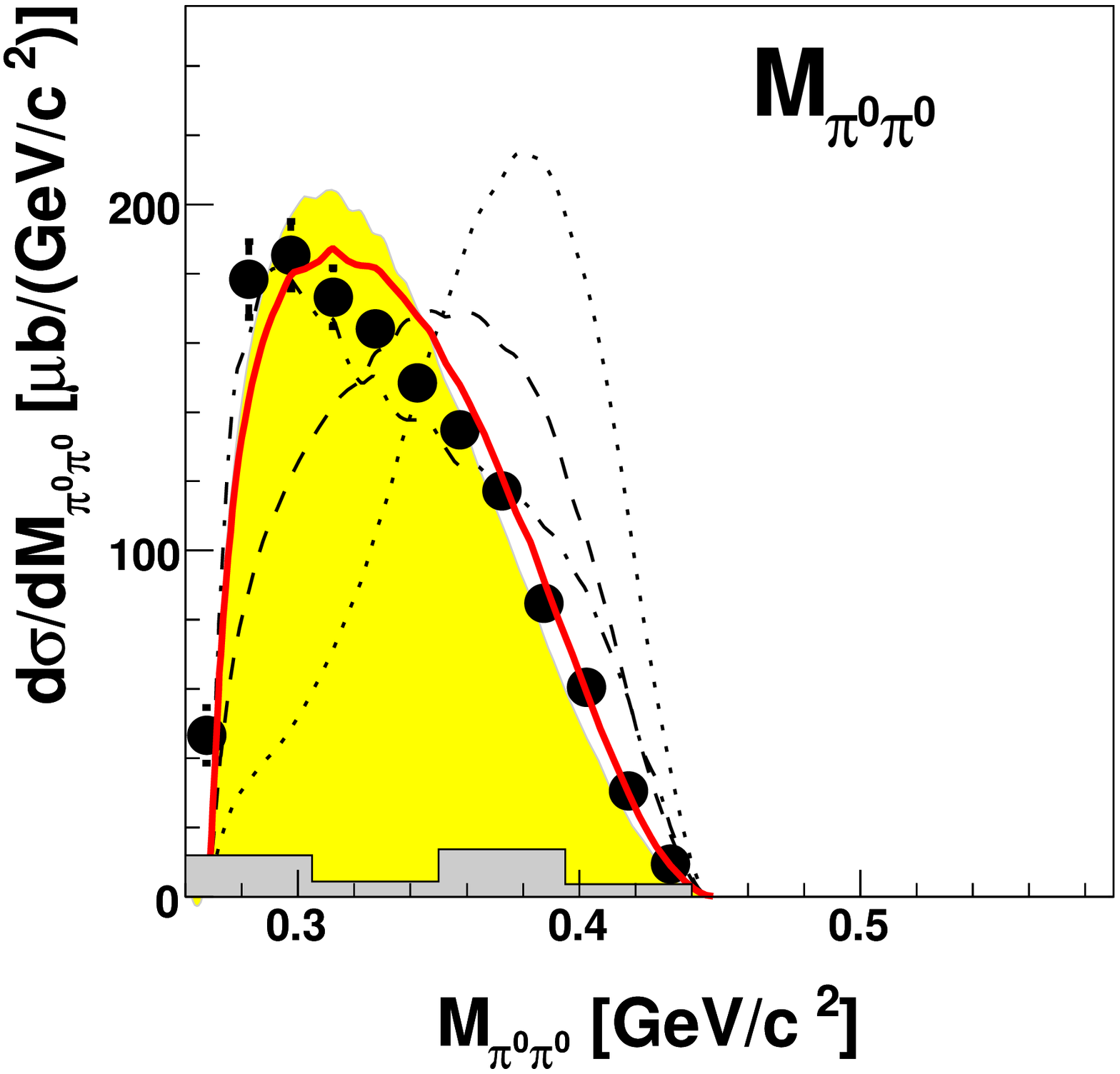}
\includegraphics[width=0.23\textwidth]{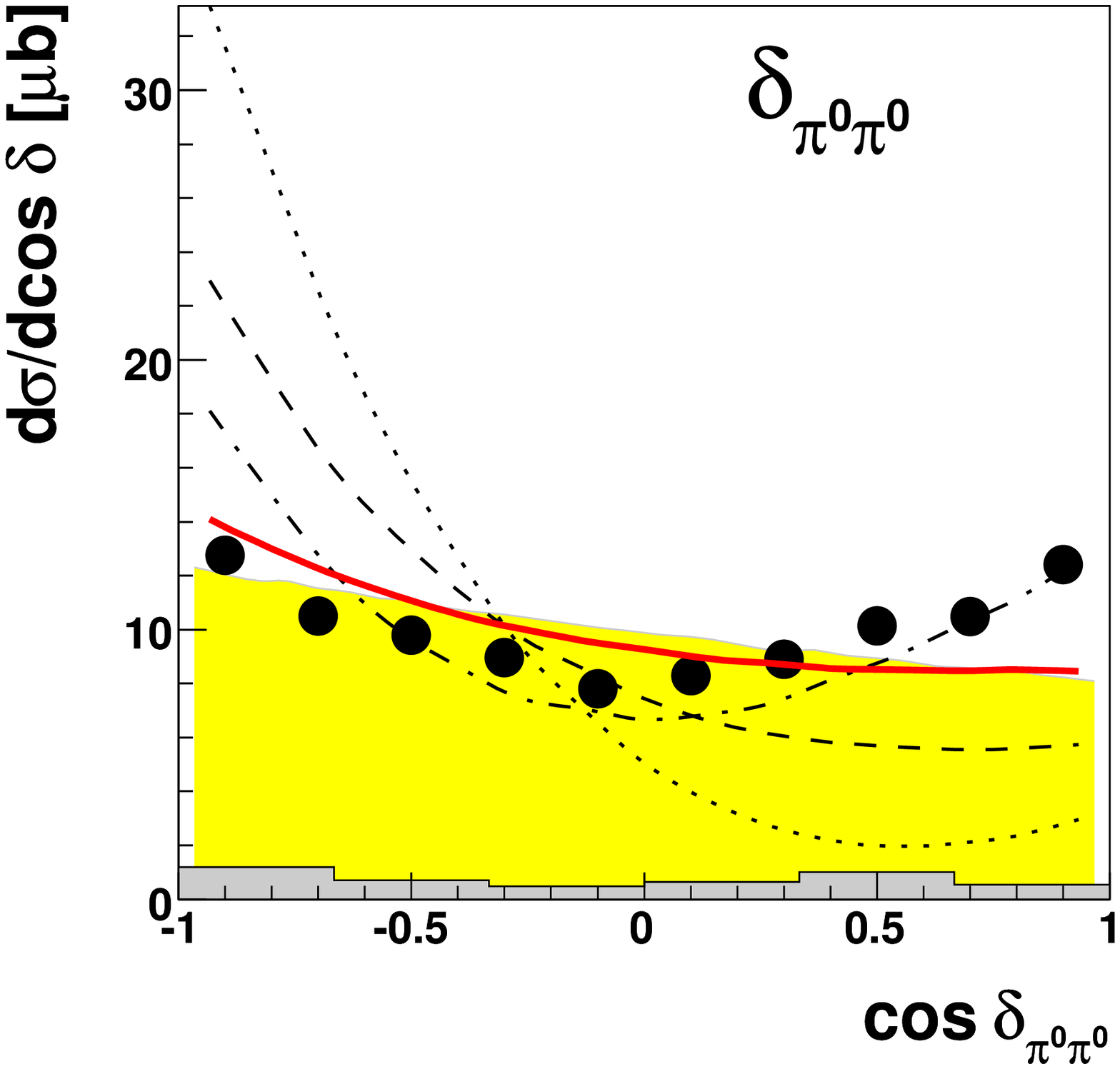}

\includegraphics[width=0.23\textwidth]{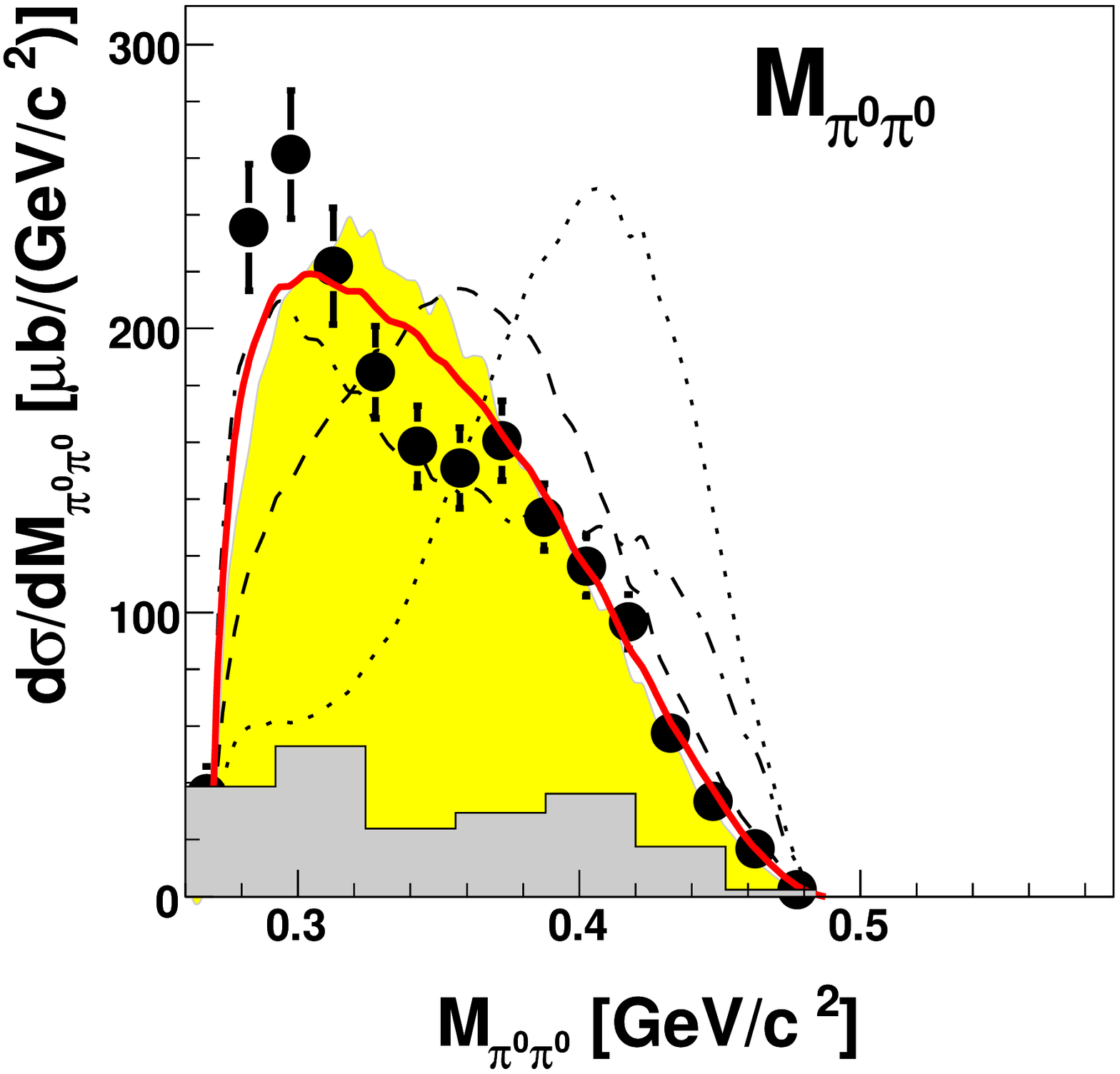}
\includegraphics[width=0.23\textwidth]{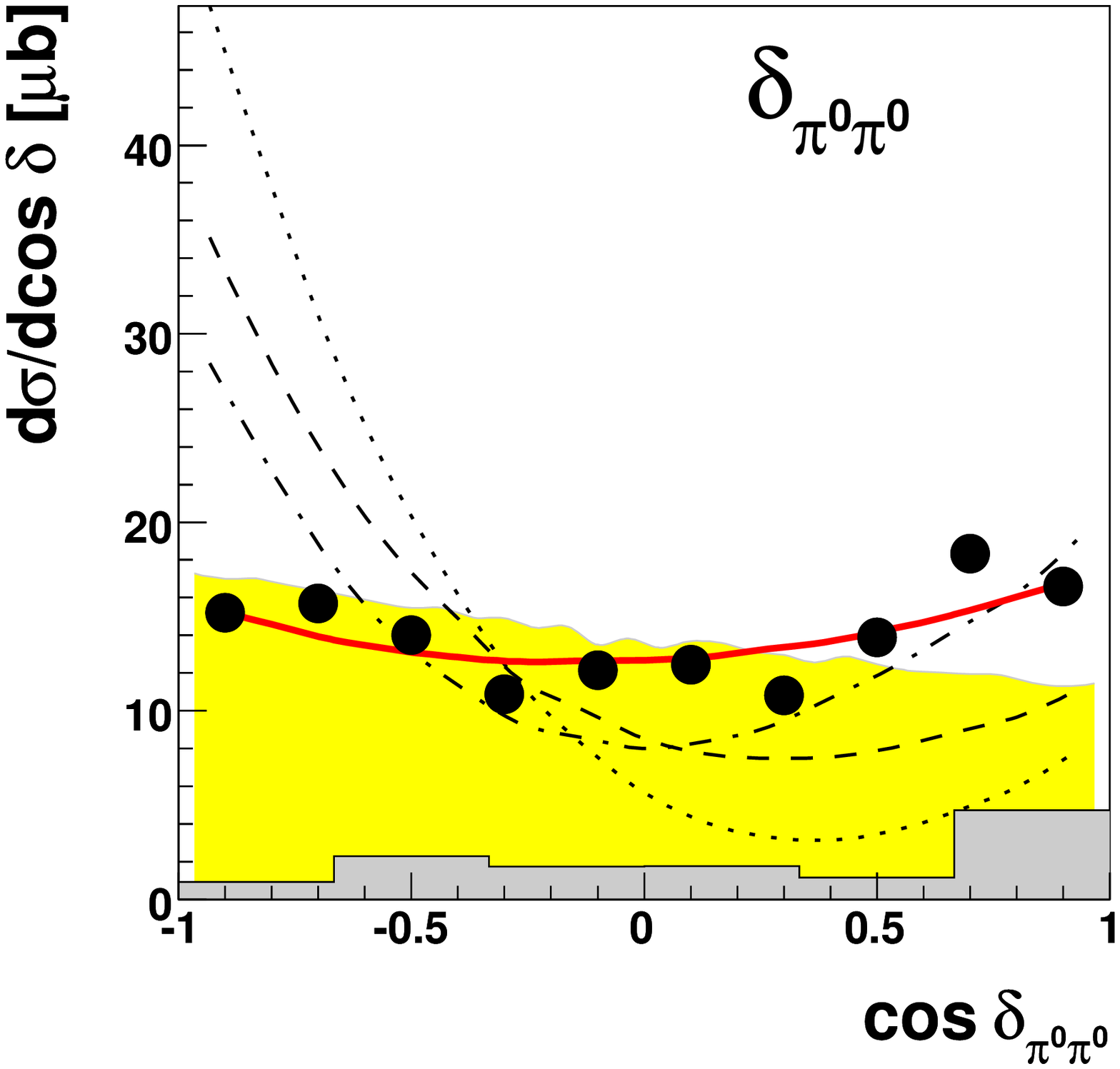}

\includegraphics[width=0.23\textwidth]{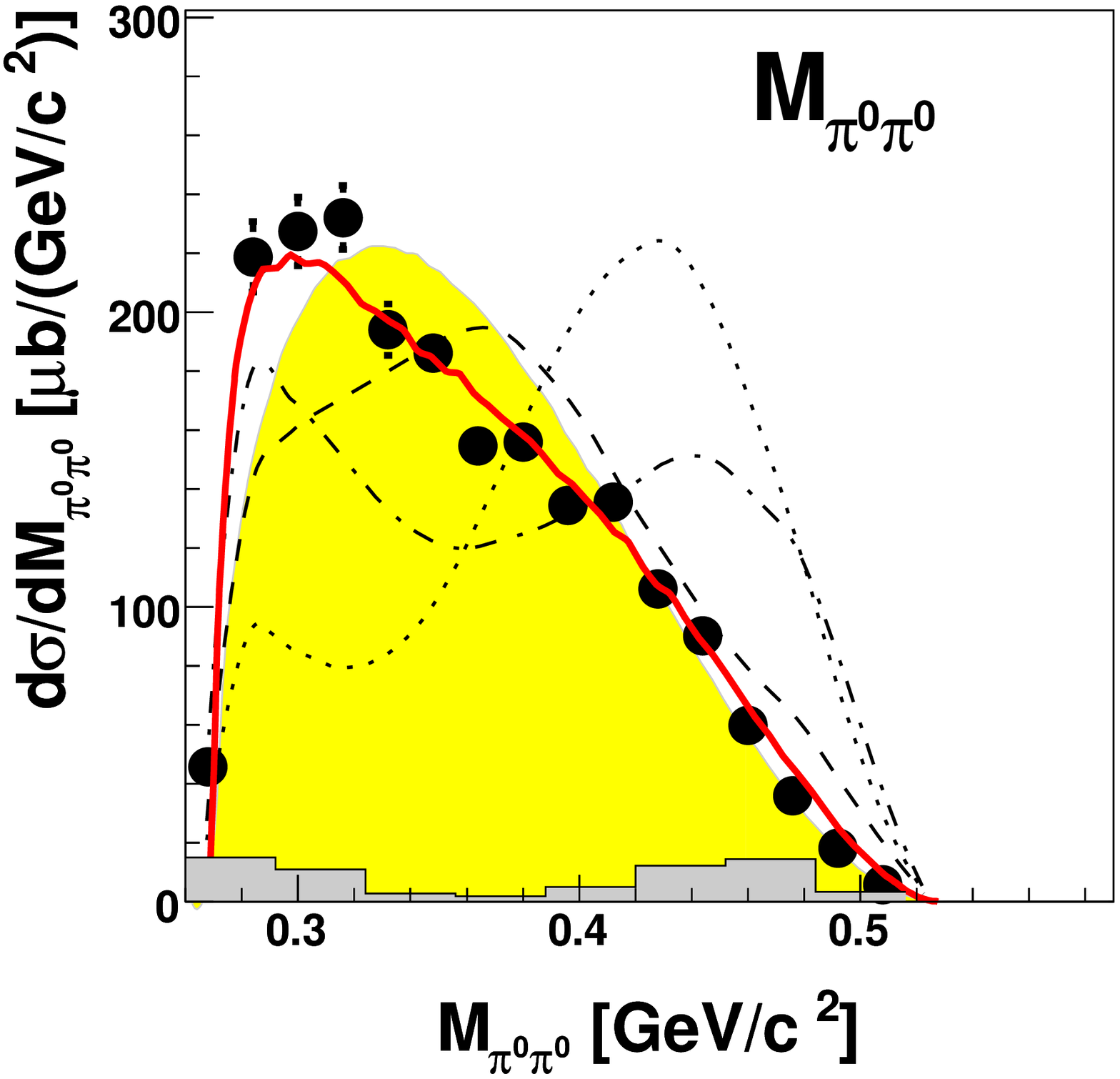}
\includegraphics[width=0.23\textwidth]{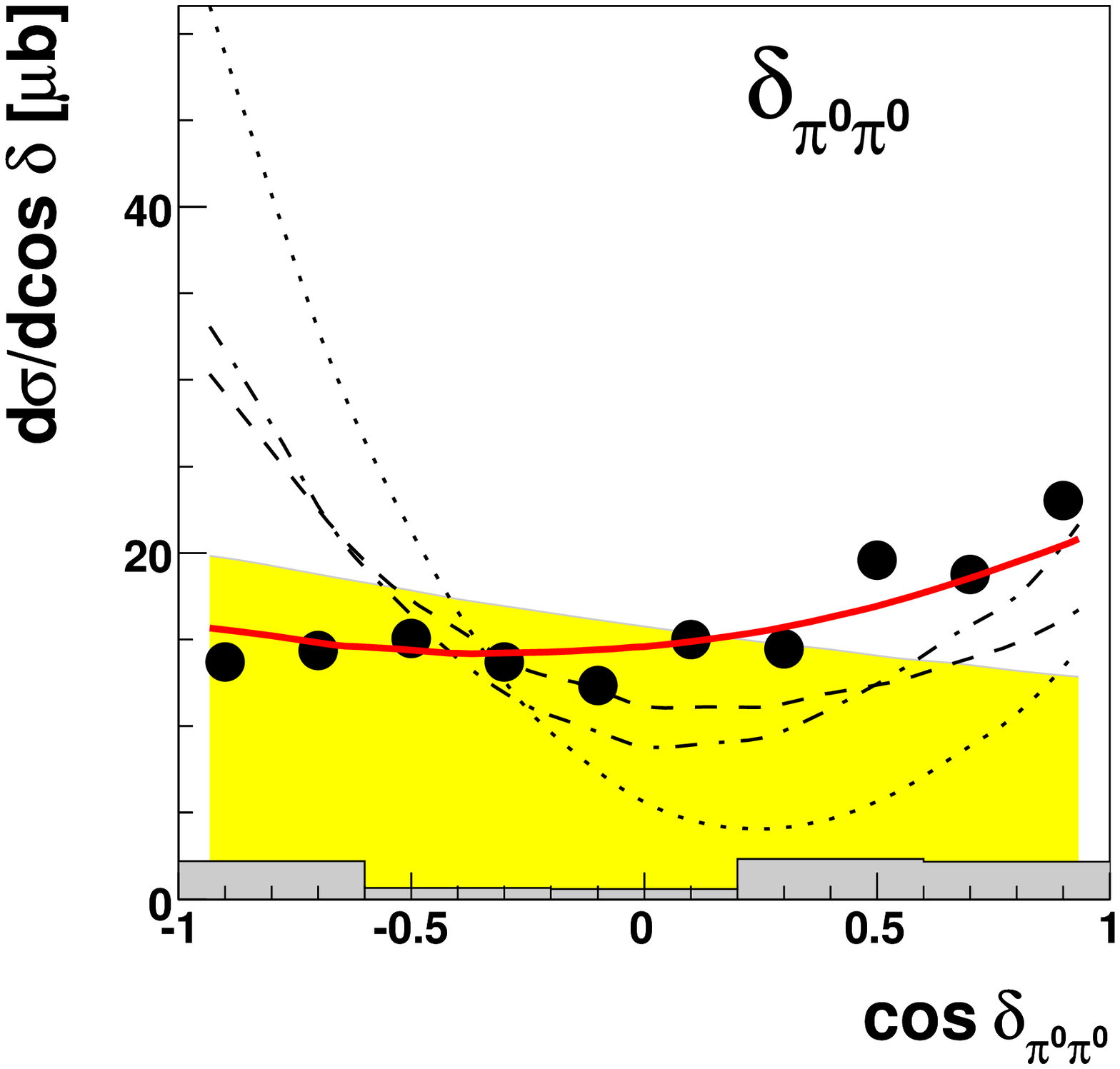}

\includegraphics[width=0.23\textwidth]{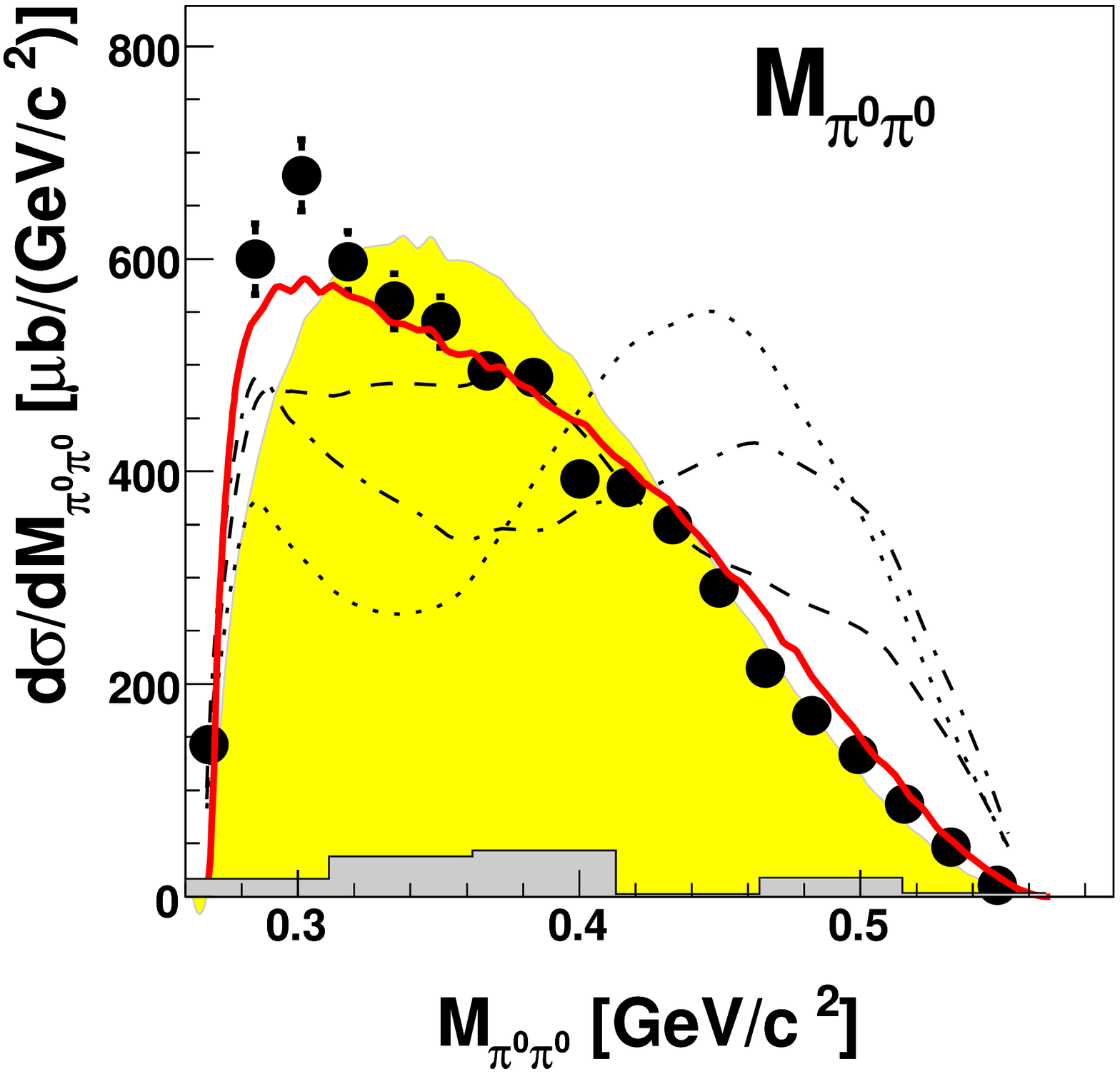}
\includegraphics[width=0.23\textwidth]{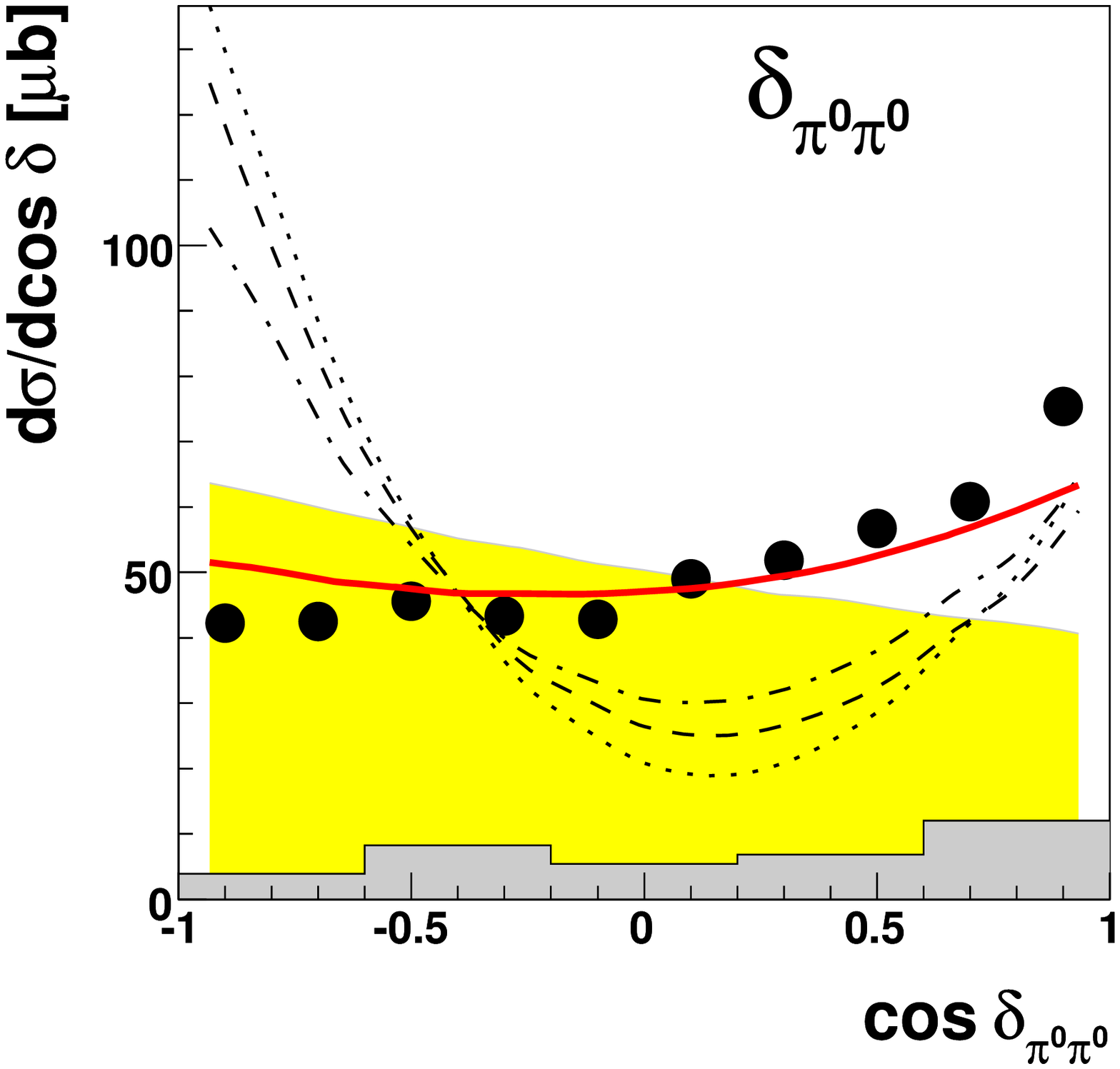}

\caption{ 
   Distribution of the $\pi^0\pi^0$ invariant mass $M_{\pi^0\pi^0}$  ({\bf
     left}) and the
   $\pi^0\pi^0$ opening angle $\delta_{\pi^0\pi^0}$  ({\bf right}) for the $pp
   \to pp\pi^0\pi^0$ reaction at beam energies $T_p$ = 1.0, 1.1, 1.2 and
   1.3 GeV (from {\bf top} to {\bf bottom}). Solid dots represent the
   experimental results of this work. The light-shaded areas denote phase space
   distributions and dark-shaded histograms systematic uncertainties. The
   dotted lines show the original calculation of 
   Ref. \cite{alv}. The dashed, dash-dotted and solid lines are calculations
   with modifications described in the text. All calculations are normalized
   in area to the data. 
}
\label{fig1}
\end{center}
\end{figure}

\begin{figure}
\begin{center}

\includegraphics[width=0.23\textwidth]{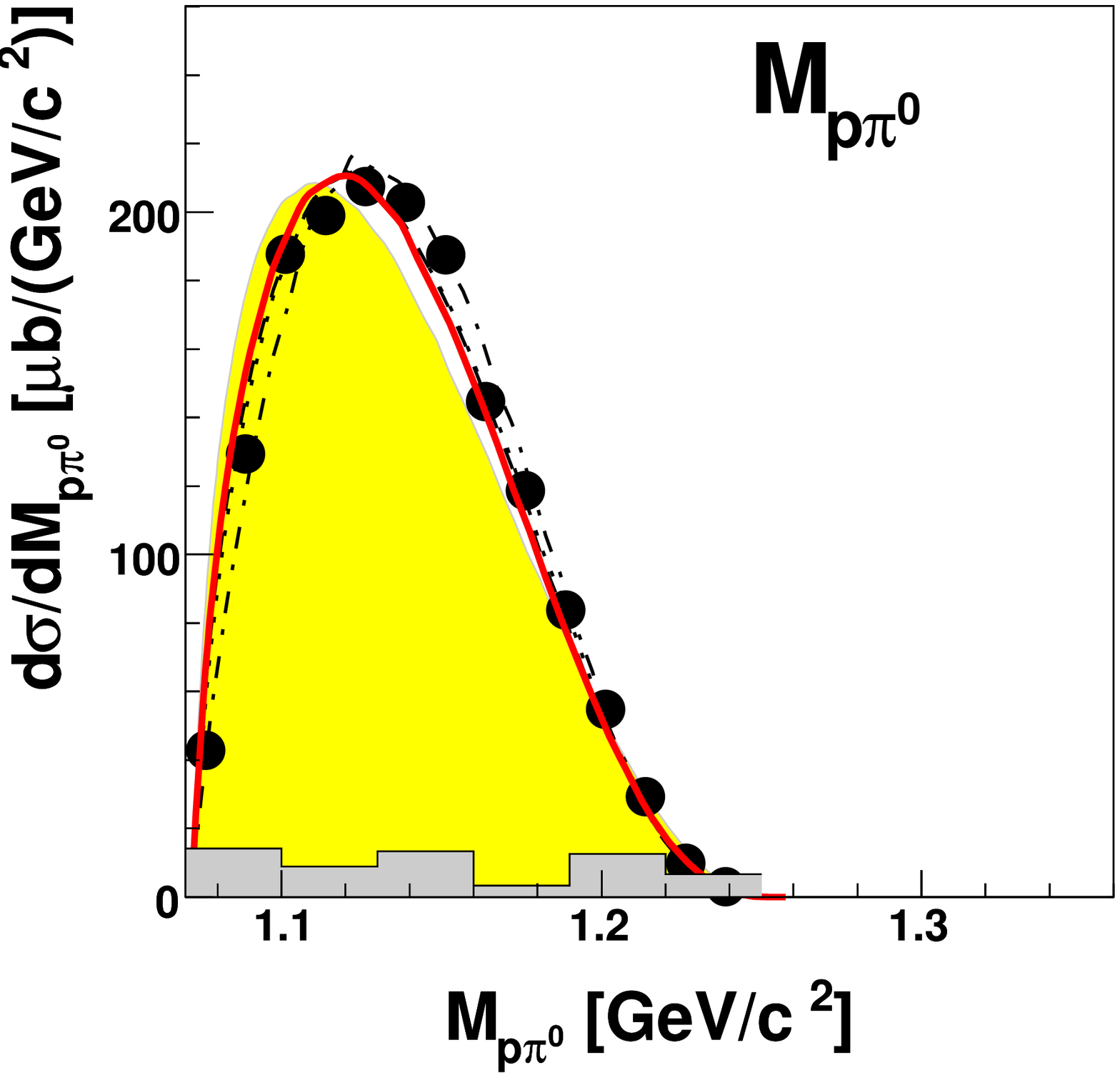}
\includegraphics[width=0.23\textwidth]{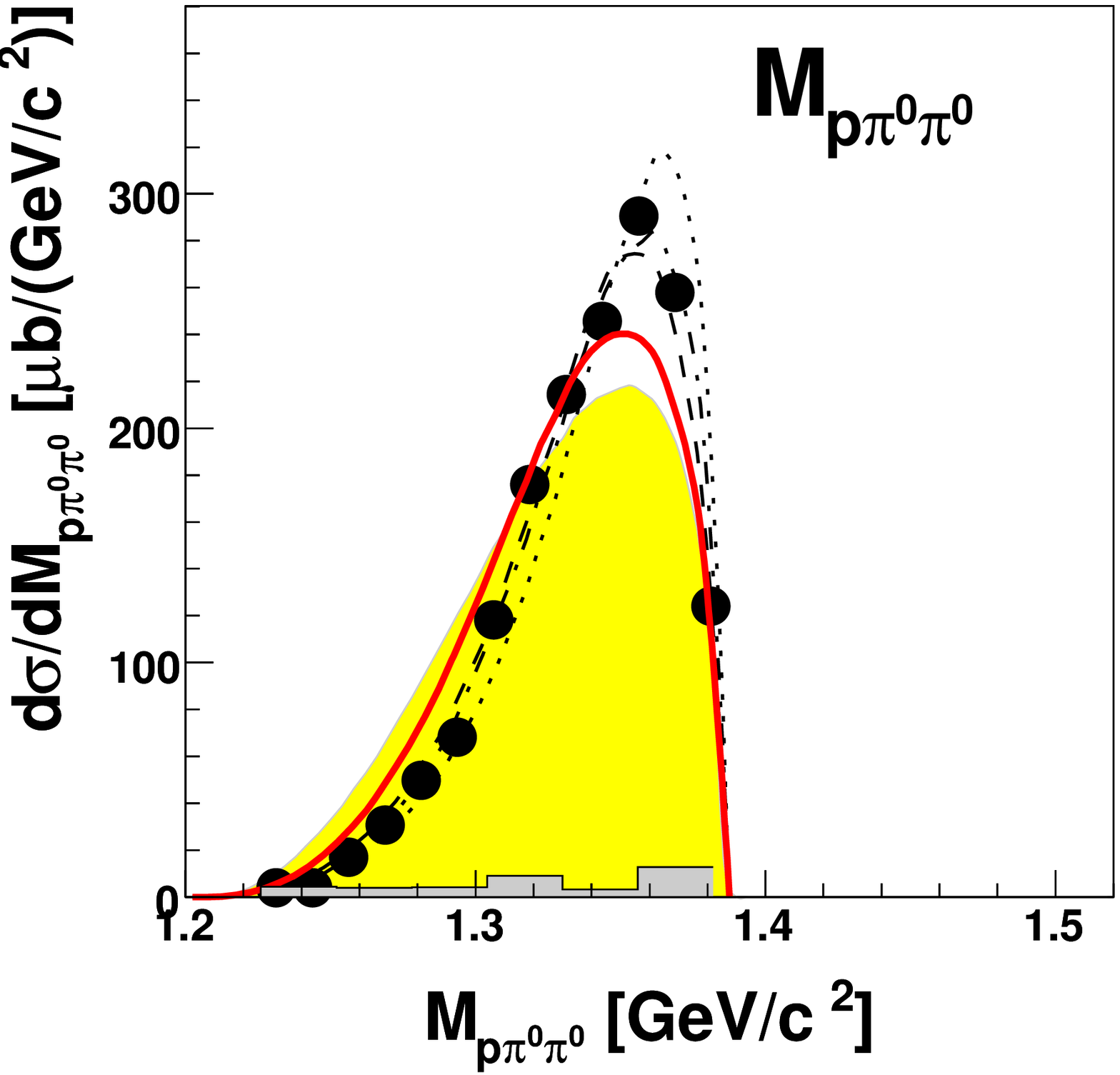}

\includegraphics[width=0.23\textwidth]{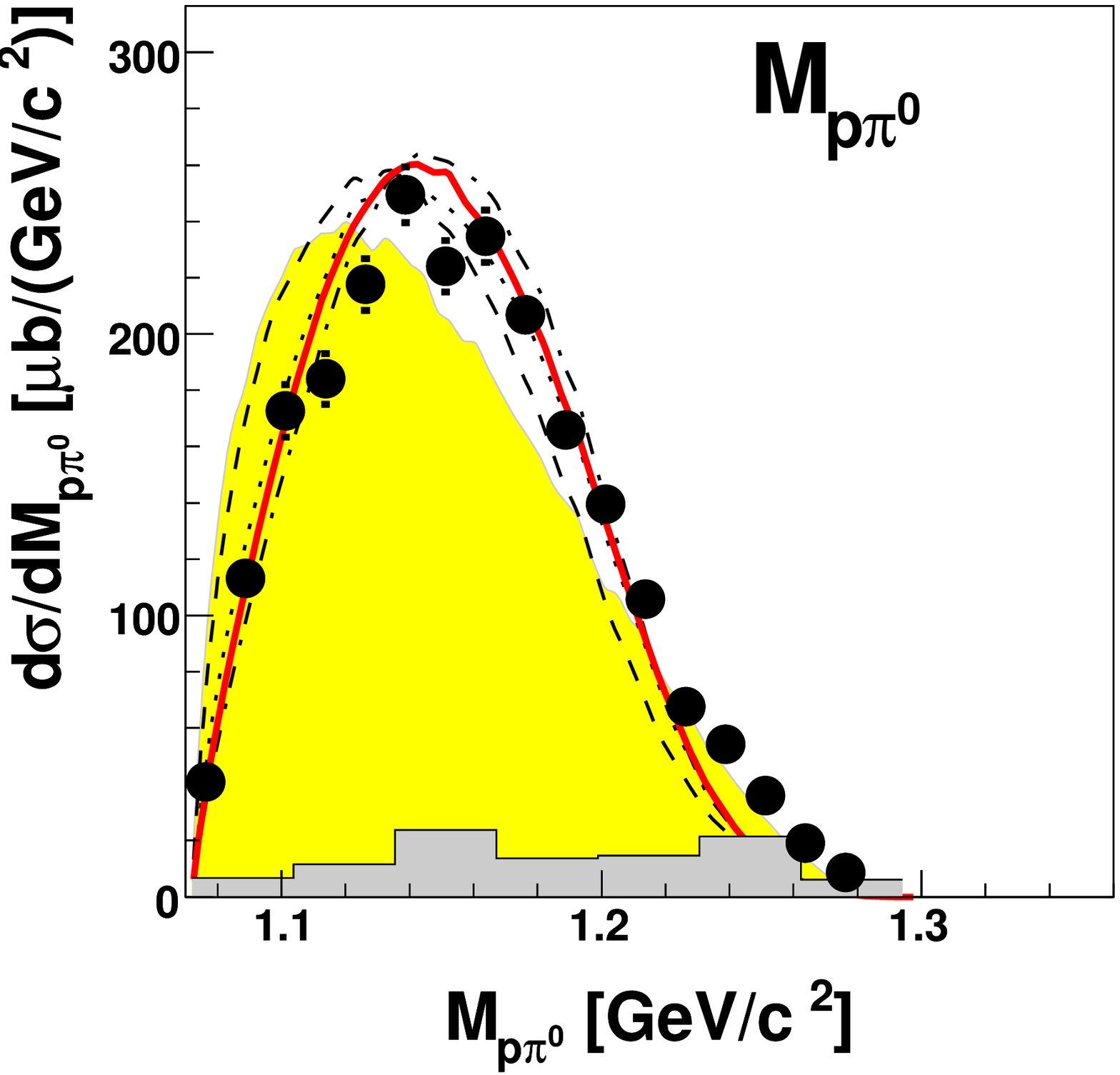}
\includegraphics[width=0.23\textwidth]{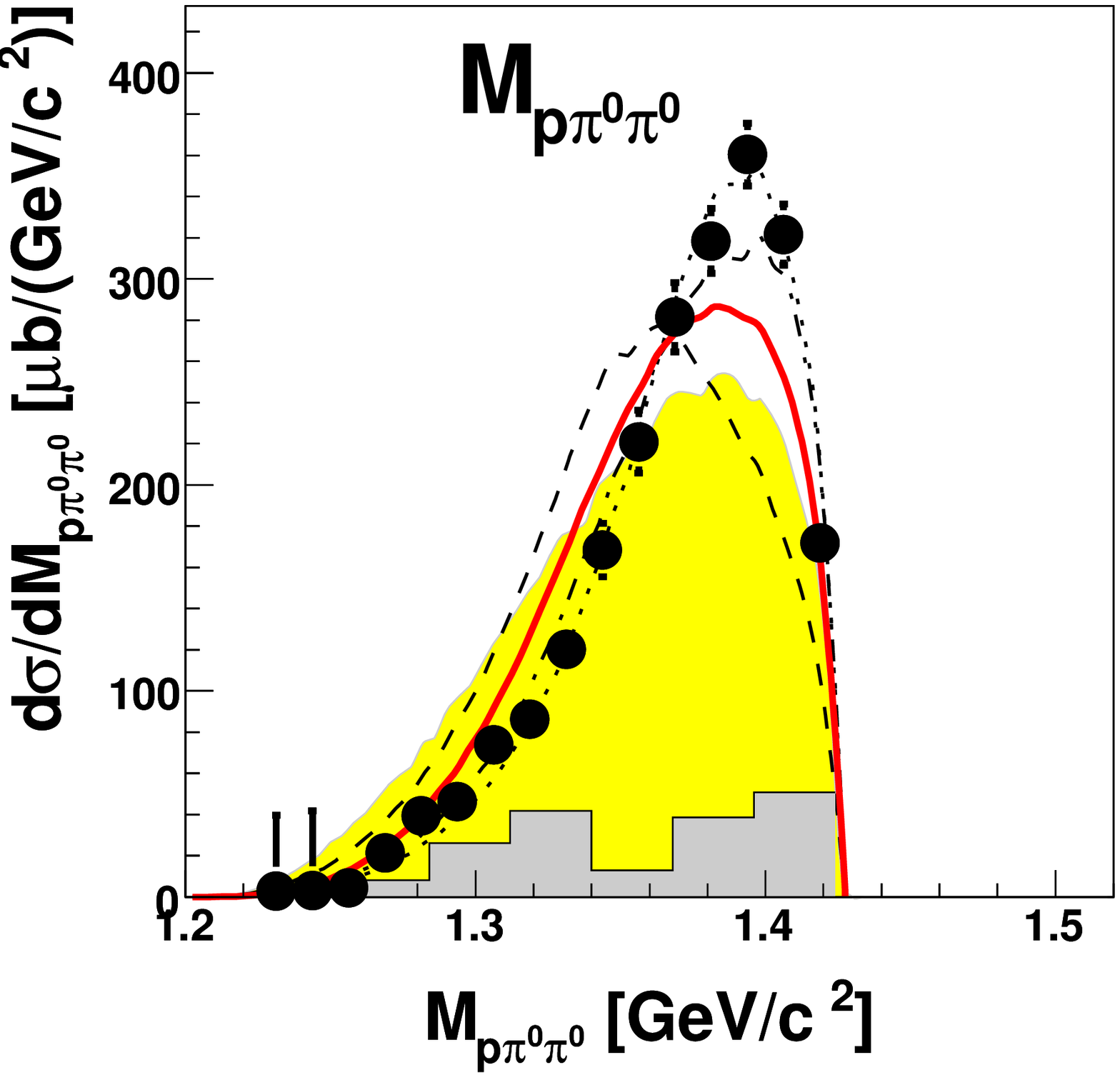}

\includegraphics[width=0.23\textwidth]{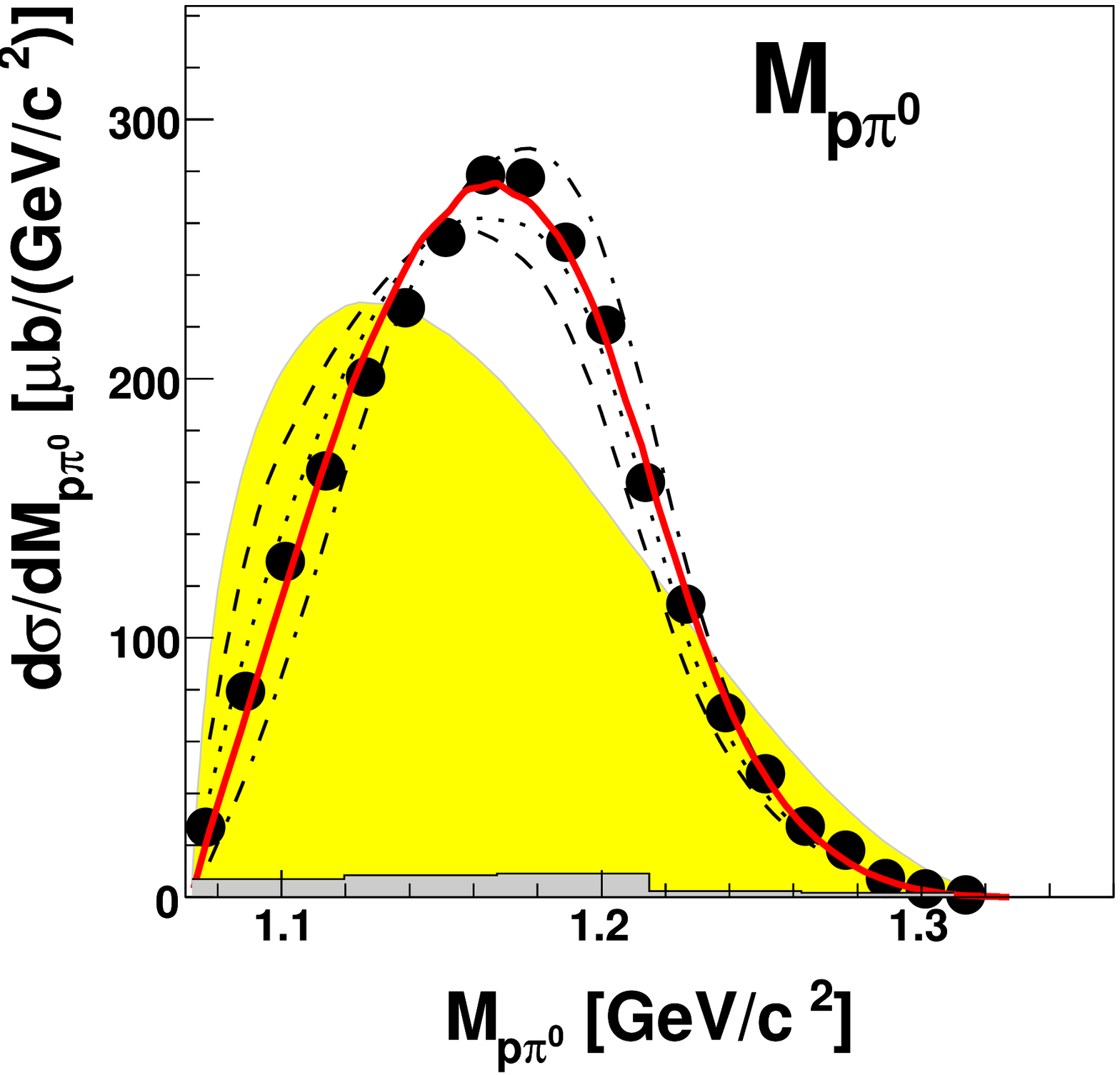}
\includegraphics[width=0.23\textwidth]{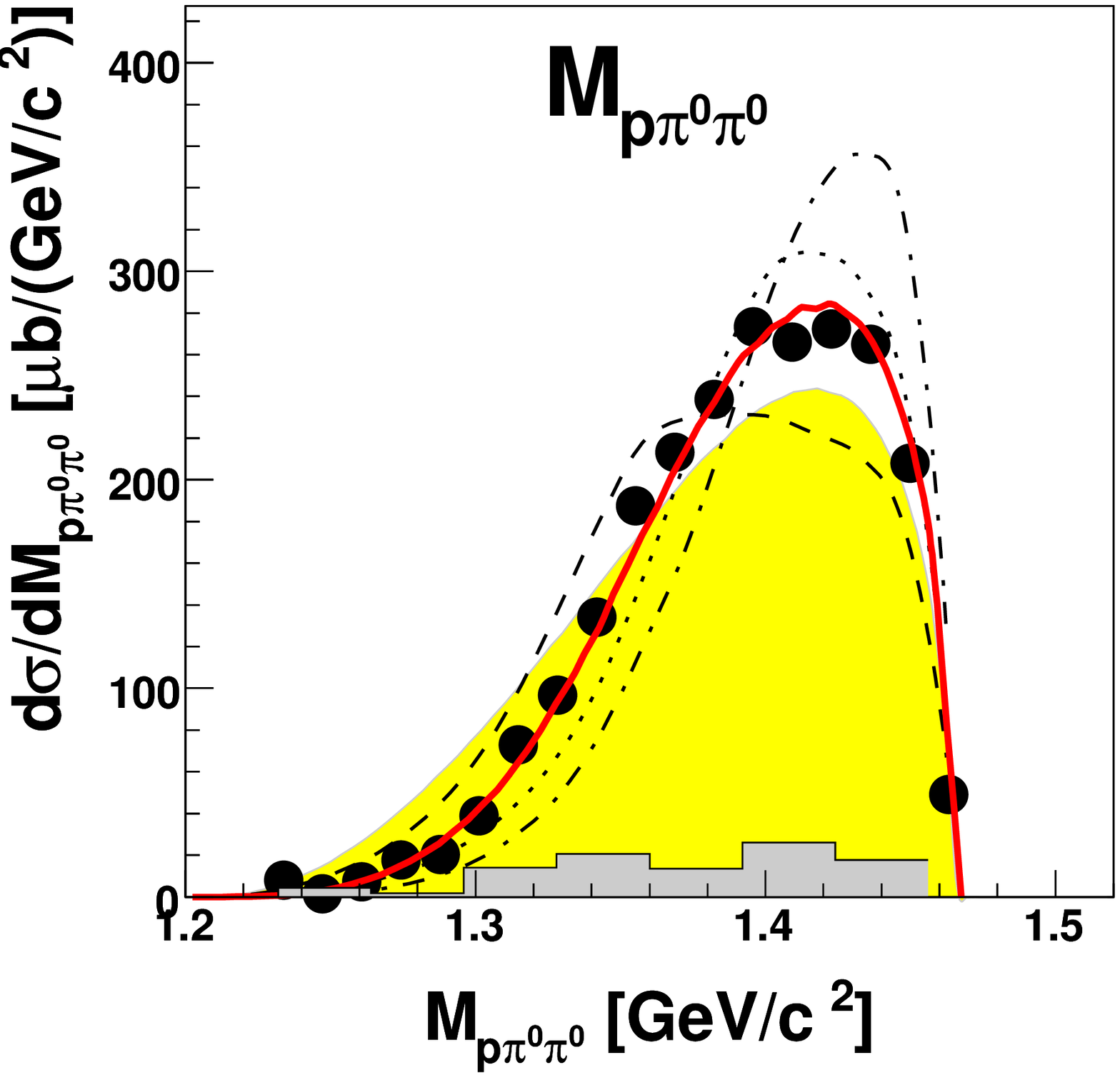}

\includegraphics[width=0.23\textwidth]{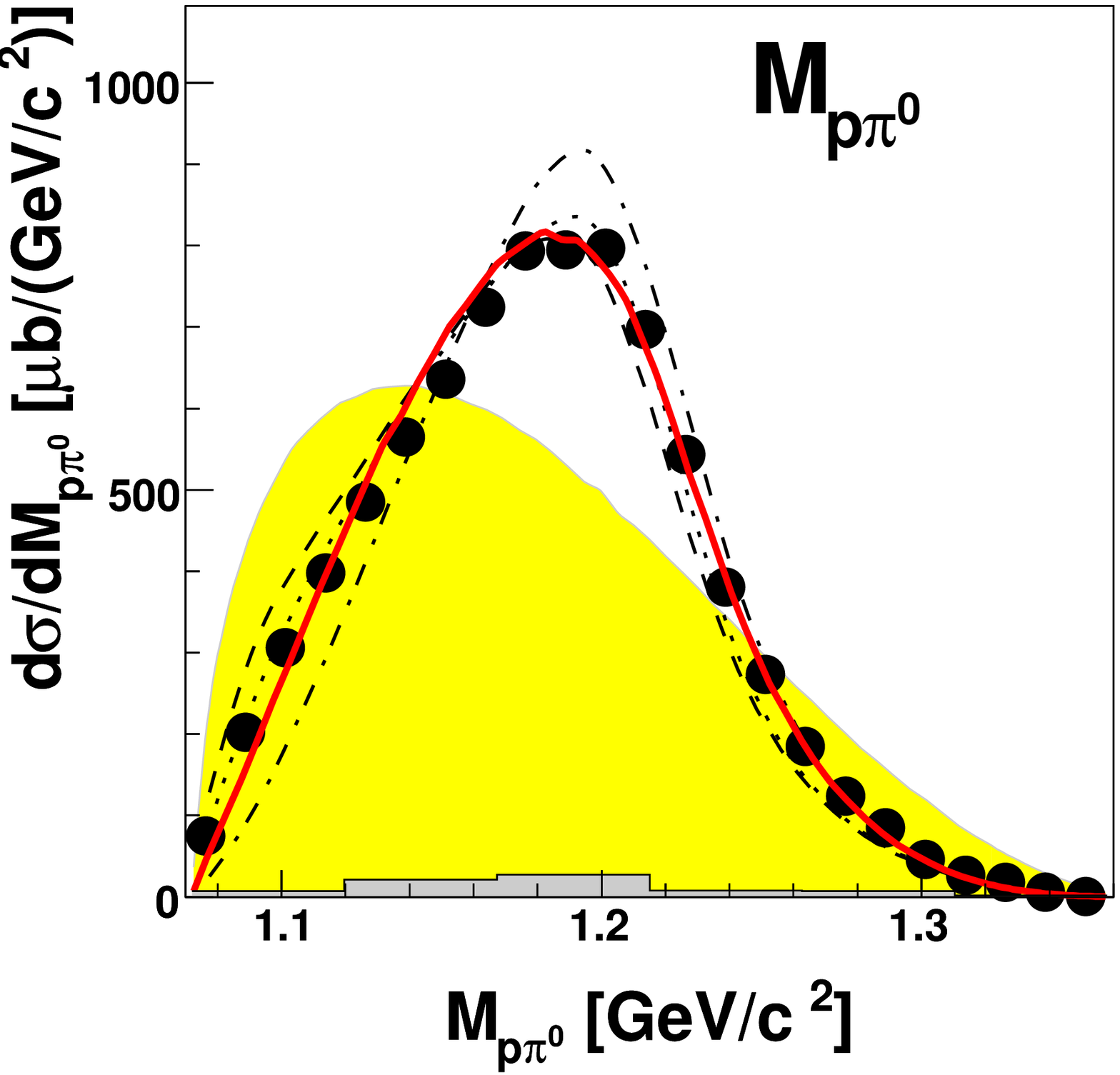}
\includegraphics[width=0.23\textwidth]{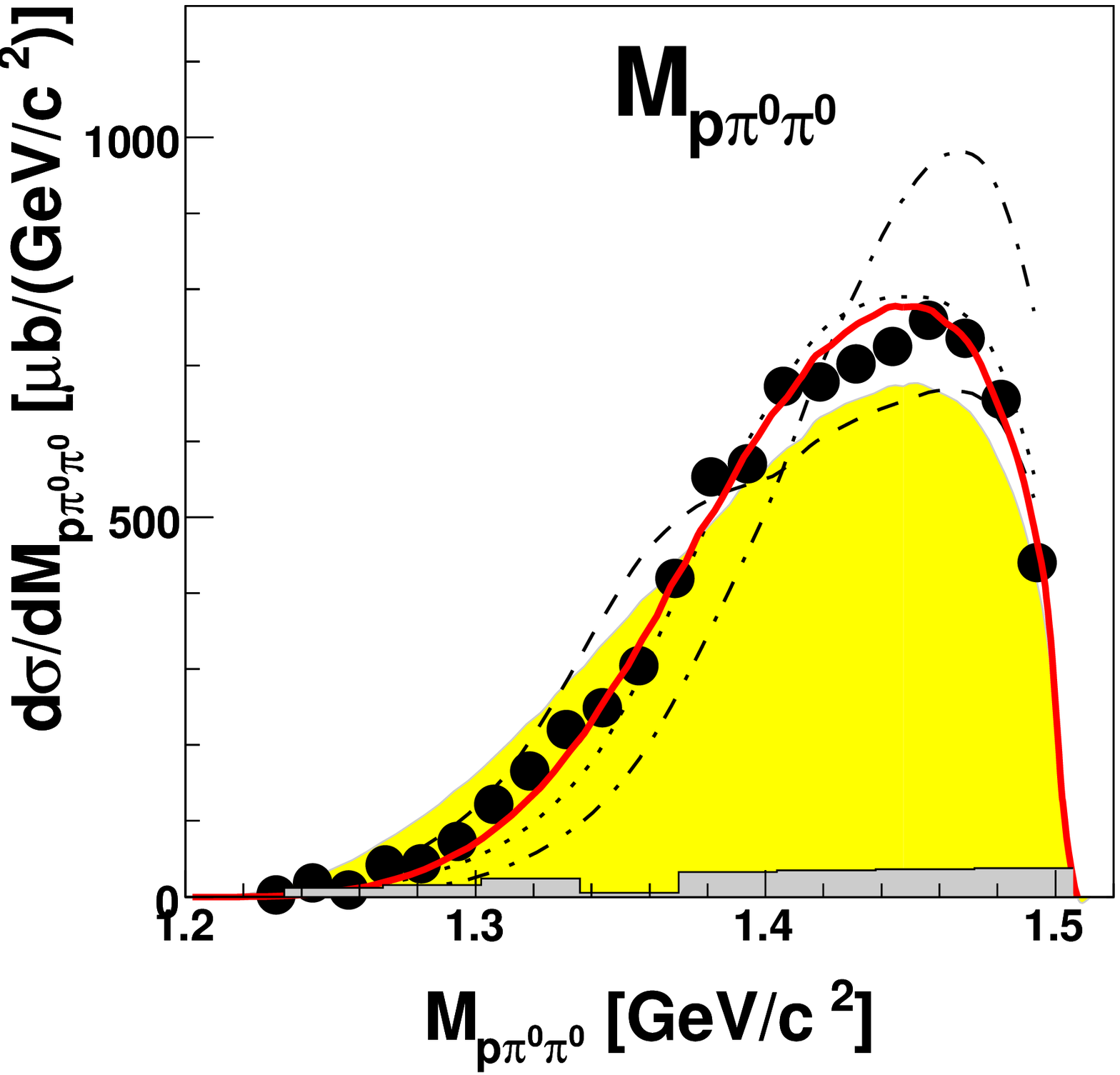}

\caption{
  Same as Fig. 2 but for the 
   distributions of the invariant masses $M_{p\pi^0}$  ({\bf
     left}) and the
   $M_{pp\pi^0}$  ({\bf right}) for the $pp
   \to pp\pi^0\pi^0$ reaction at beam energies $T_p$ = 1.0, 1.1, 1.2 and
   1.3 GeV (from {\bf top} to {\bf bottom}). 
}
\label{fig2}
\end{center}
\end{figure}

\begin{figure}
\begin{center}

\includegraphics[width=0.23\textwidth]{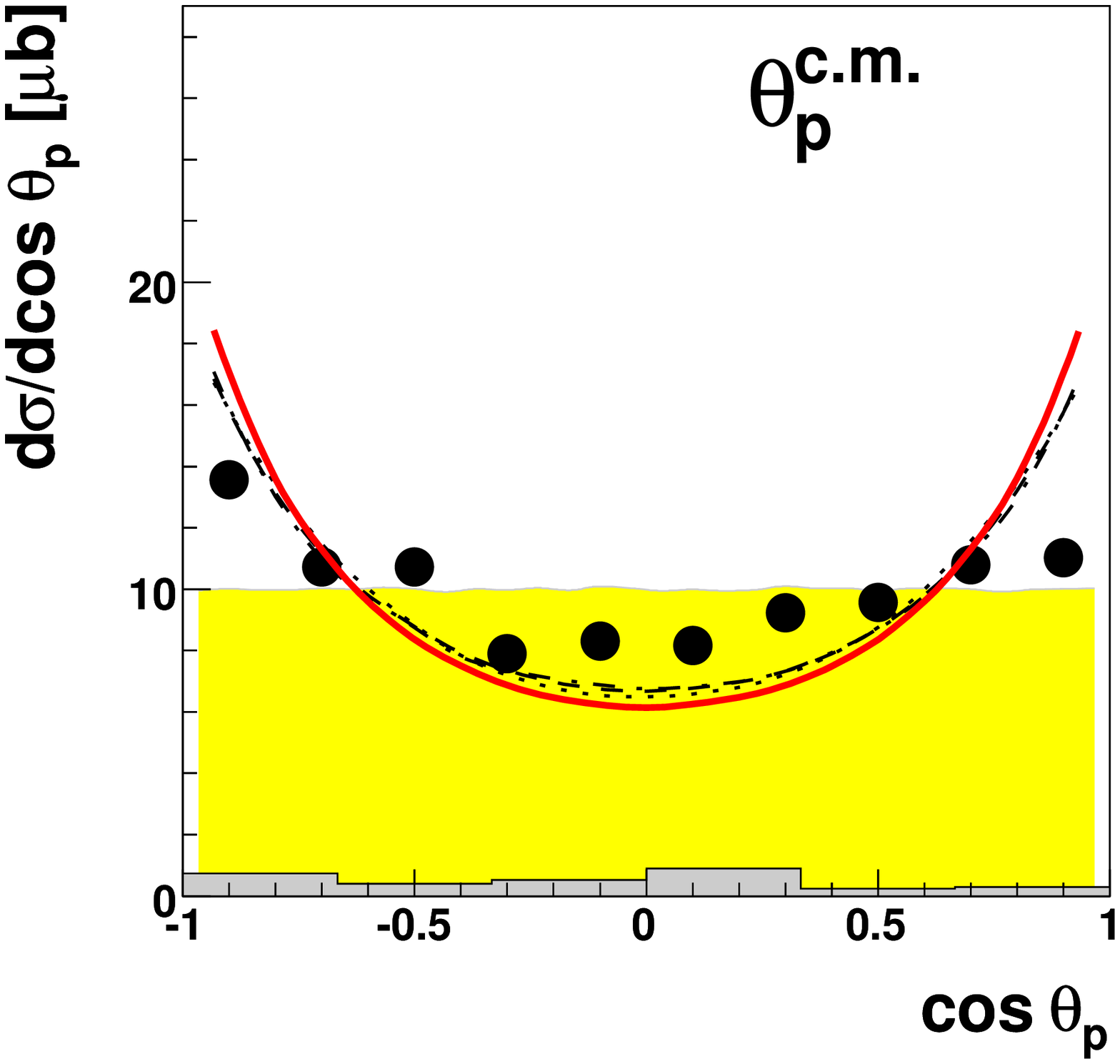}
\includegraphics[width=0.23\textwidth]{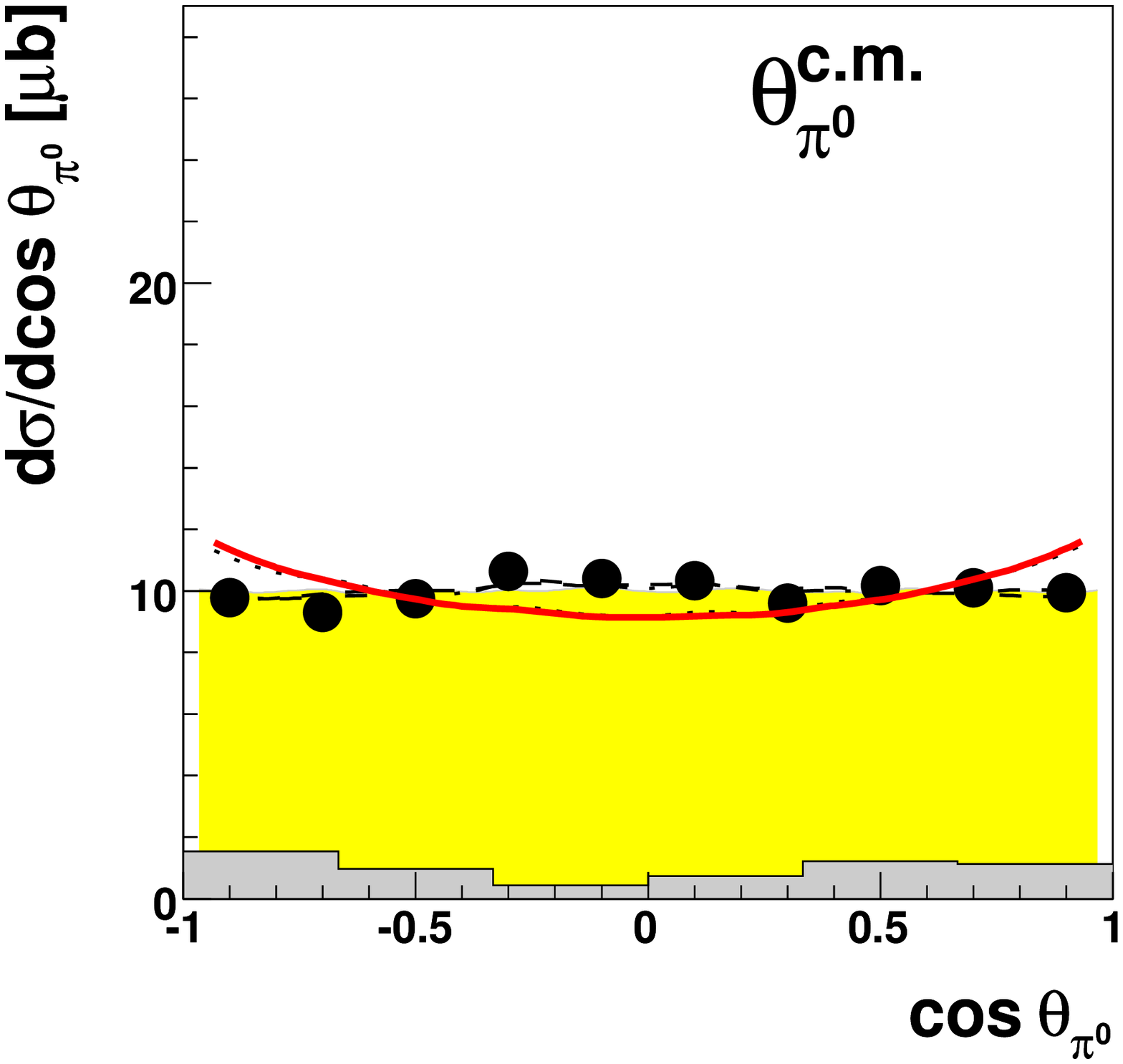}

\includegraphics[width=0.23\textwidth]{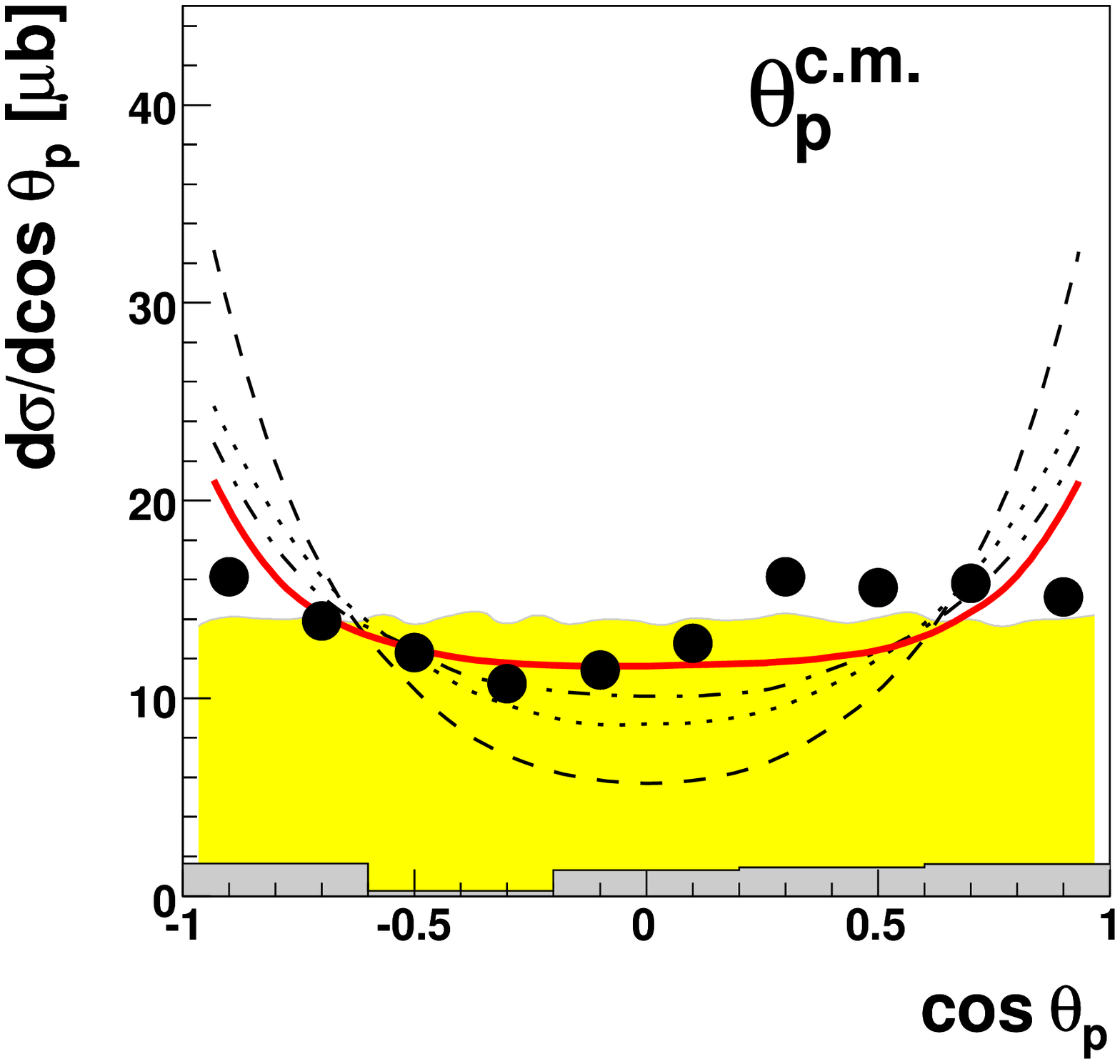}
\includegraphics[width=0.23\textwidth]{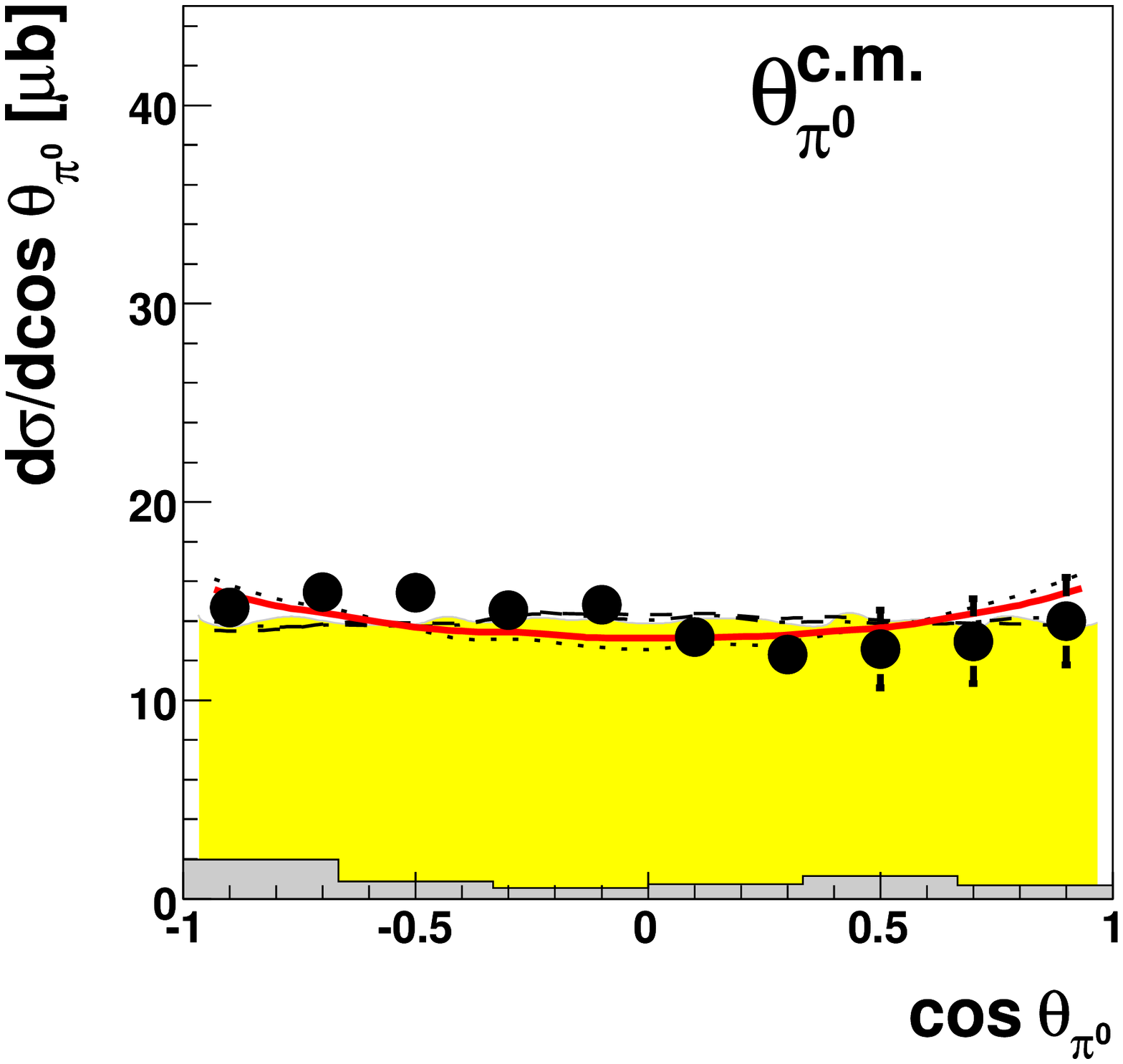}

\includegraphics[width=0.23\textwidth]{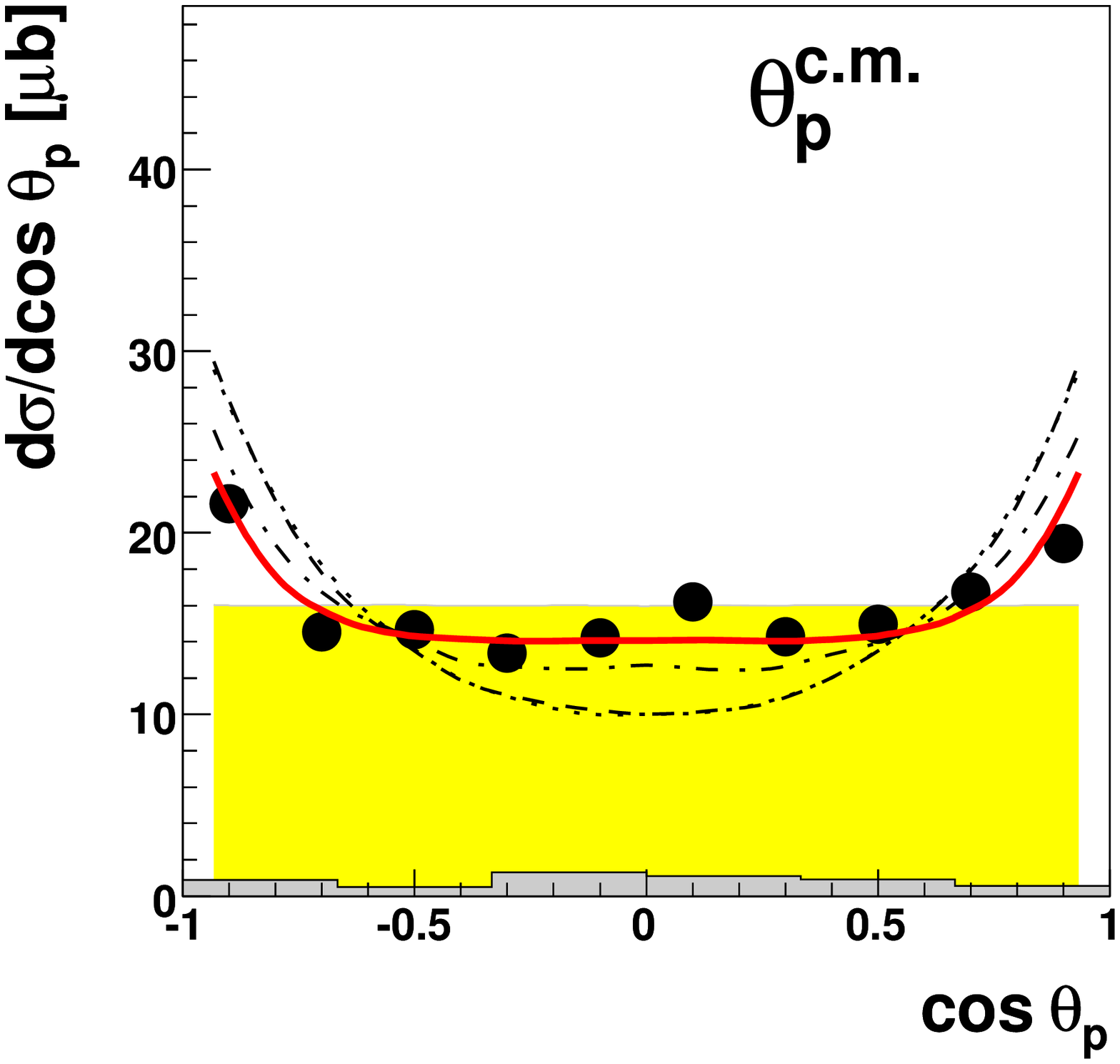}
\includegraphics[width=0.23\textwidth]{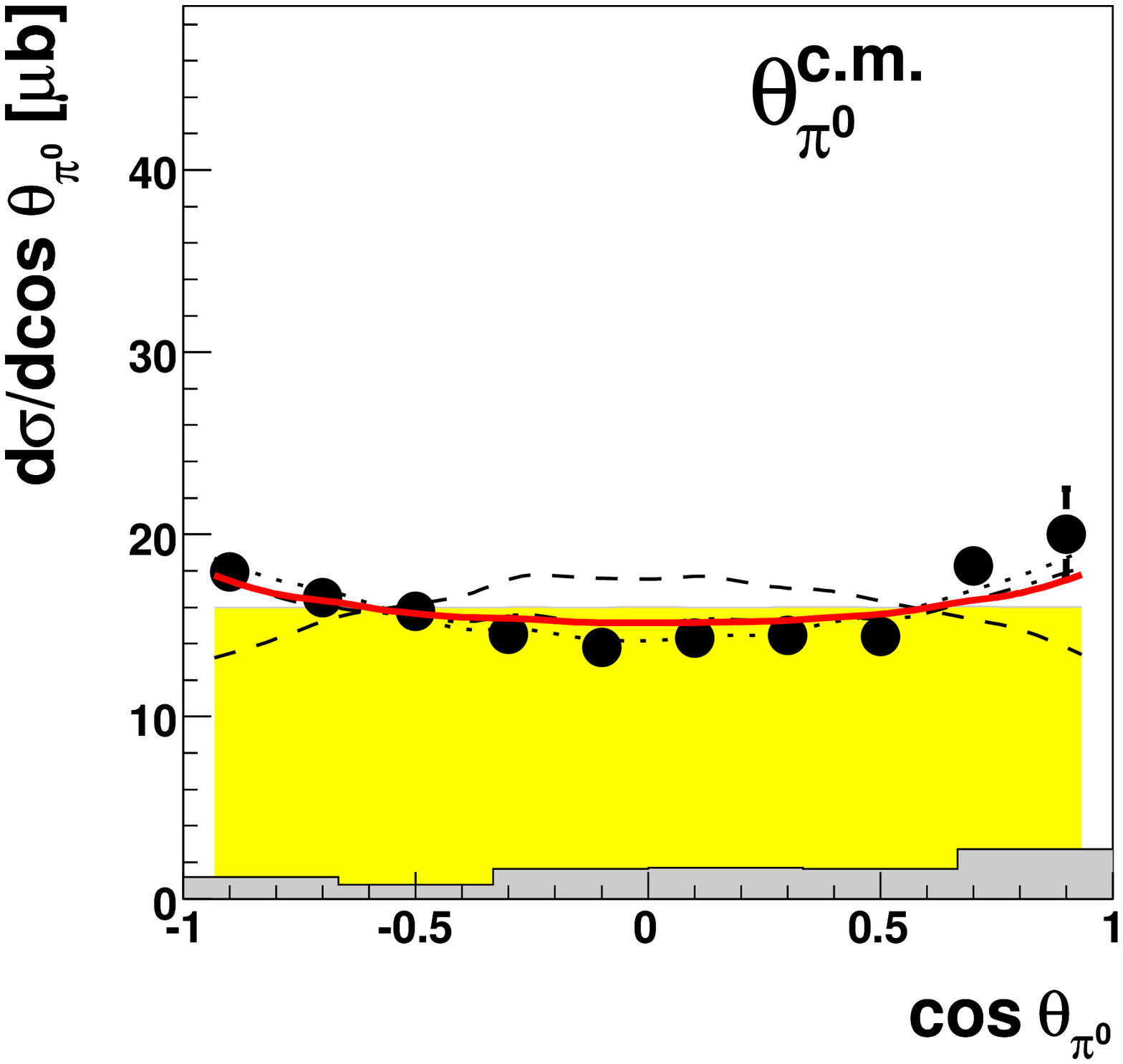}

\includegraphics[width=0.23\textwidth]{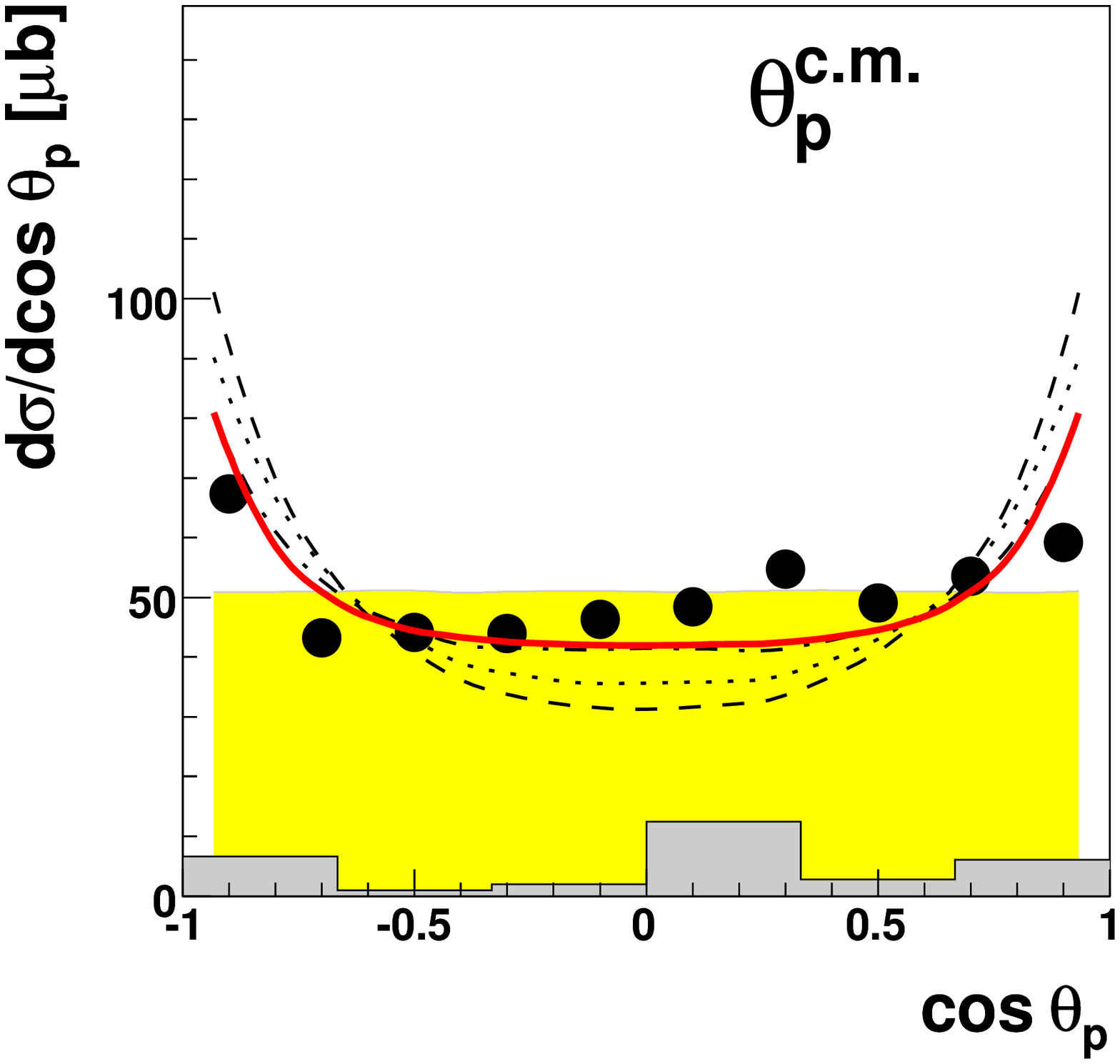}
\includegraphics[width=0.23\textwidth]{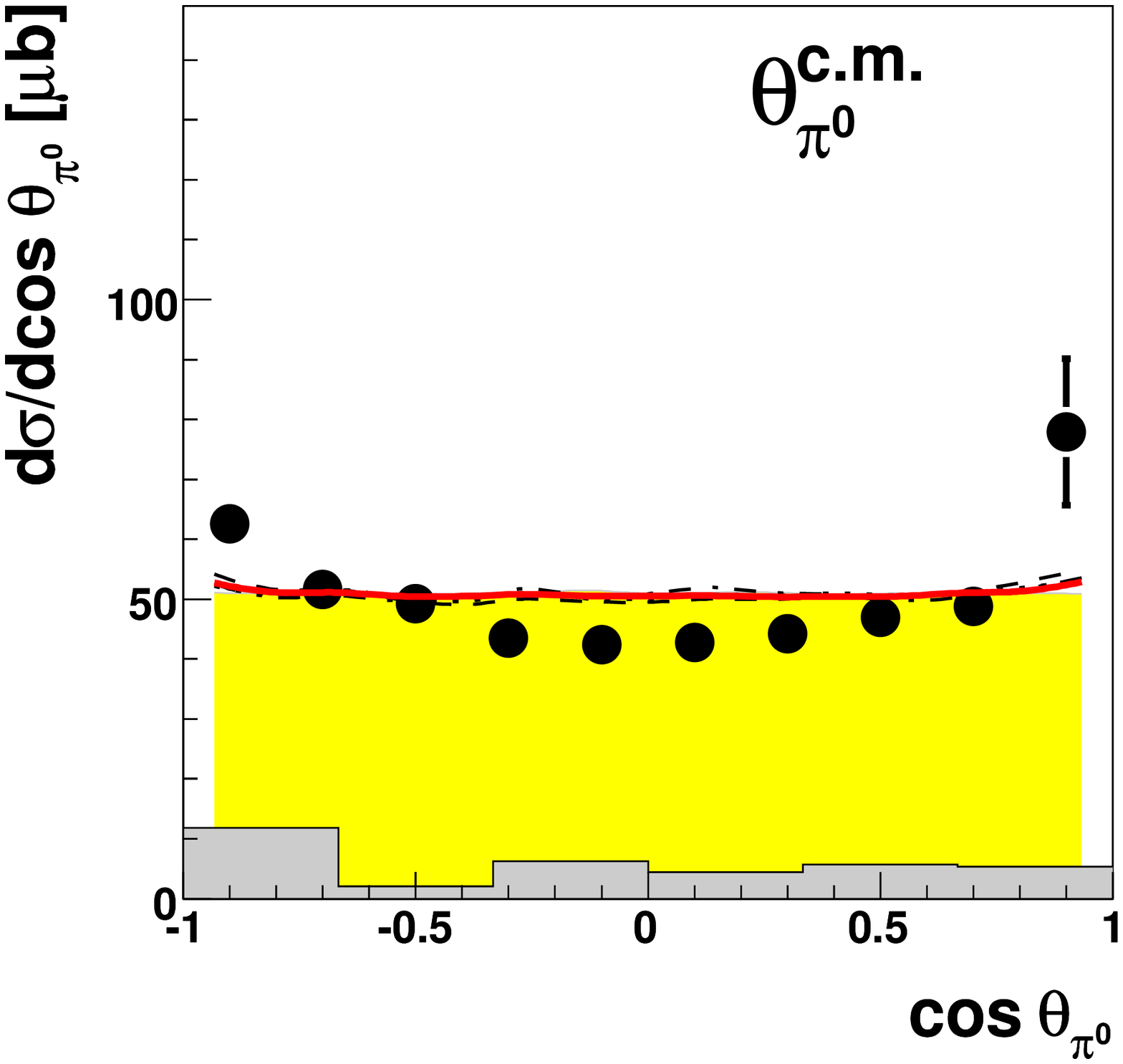}

\caption{
  Same as Fig. 2 but for the 
   distributions of the cms angles $\Theta_p$  ({\bf
     left}) and the
   $\Theta_{\pi^0}$  ({\bf right}) for the $pp
   \to pp\pi^0\pi^0$ reaction at beam energies $T_p$ = 1.0, 1.1, 1.2 and
   1.3 GeV (from {\bf top} to {\bf bottom}). 
}
\label{fig3}
\end{center}
\end{figure}

\begin{figure}
\begin{center}

\includegraphics[width=0.23\textwidth]{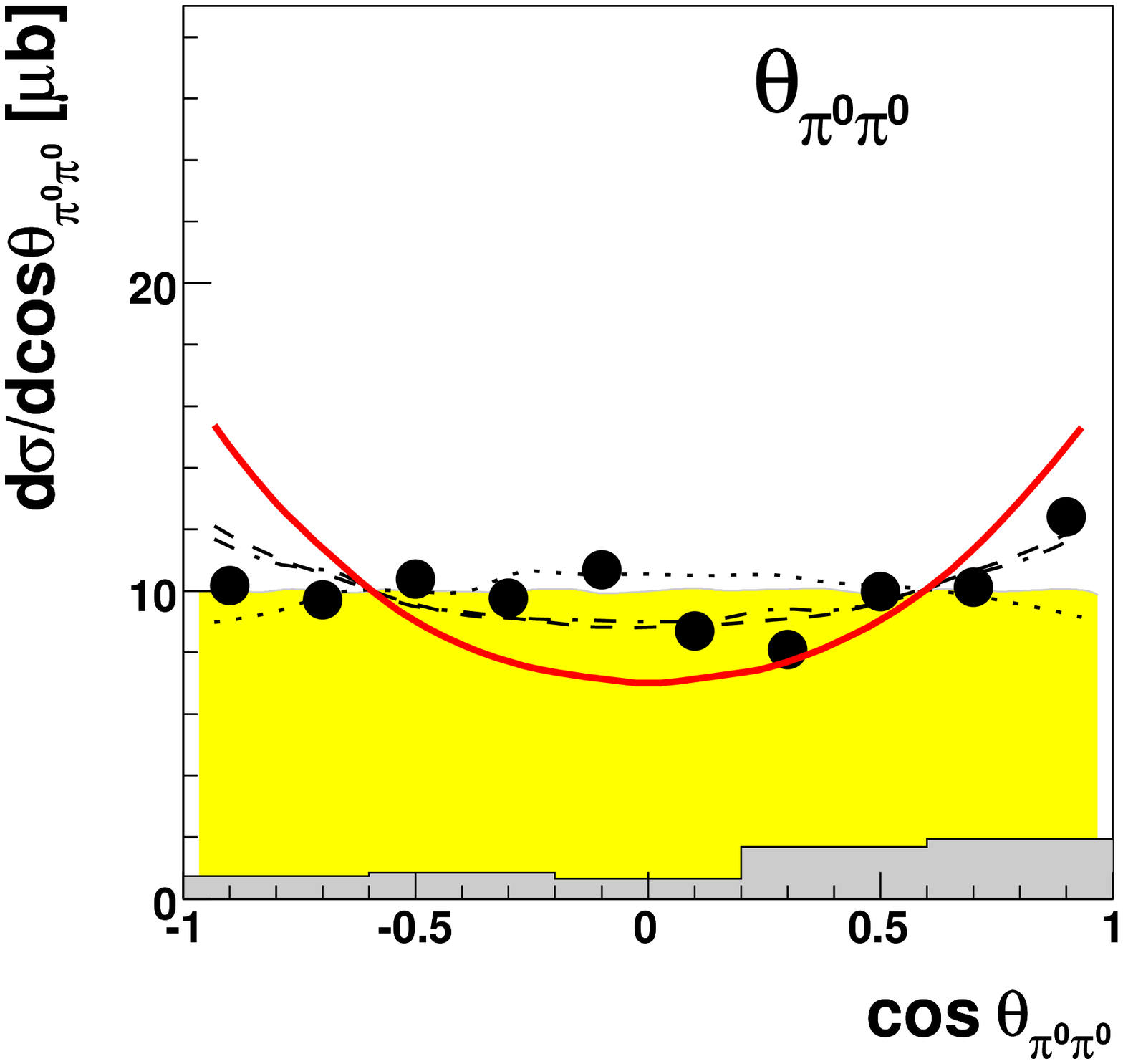}
\includegraphics[width=0.23\textwidth]{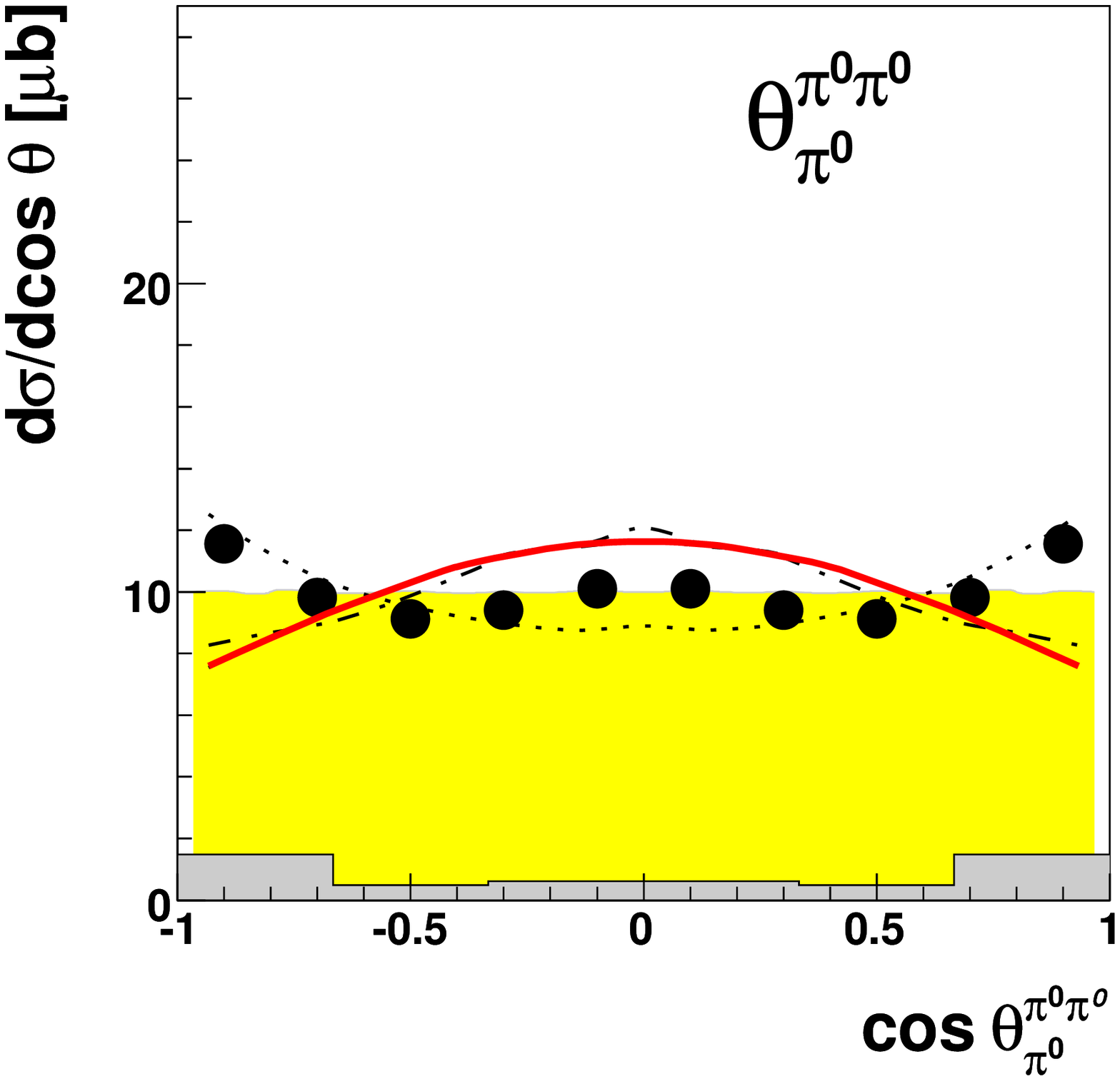}

\includegraphics[width=0.23\textwidth]{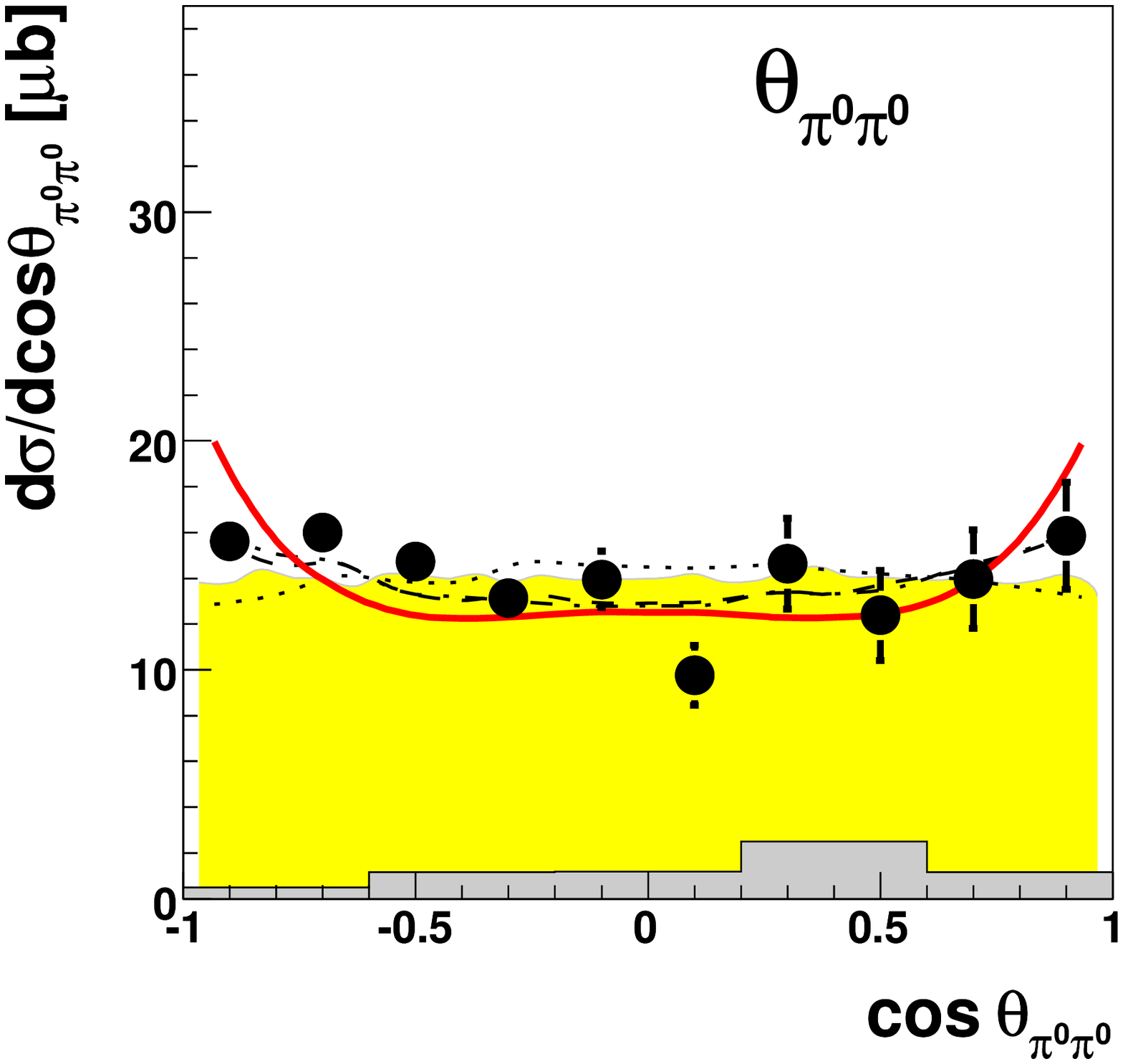}
\includegraphics[width=0.23\textwidth]{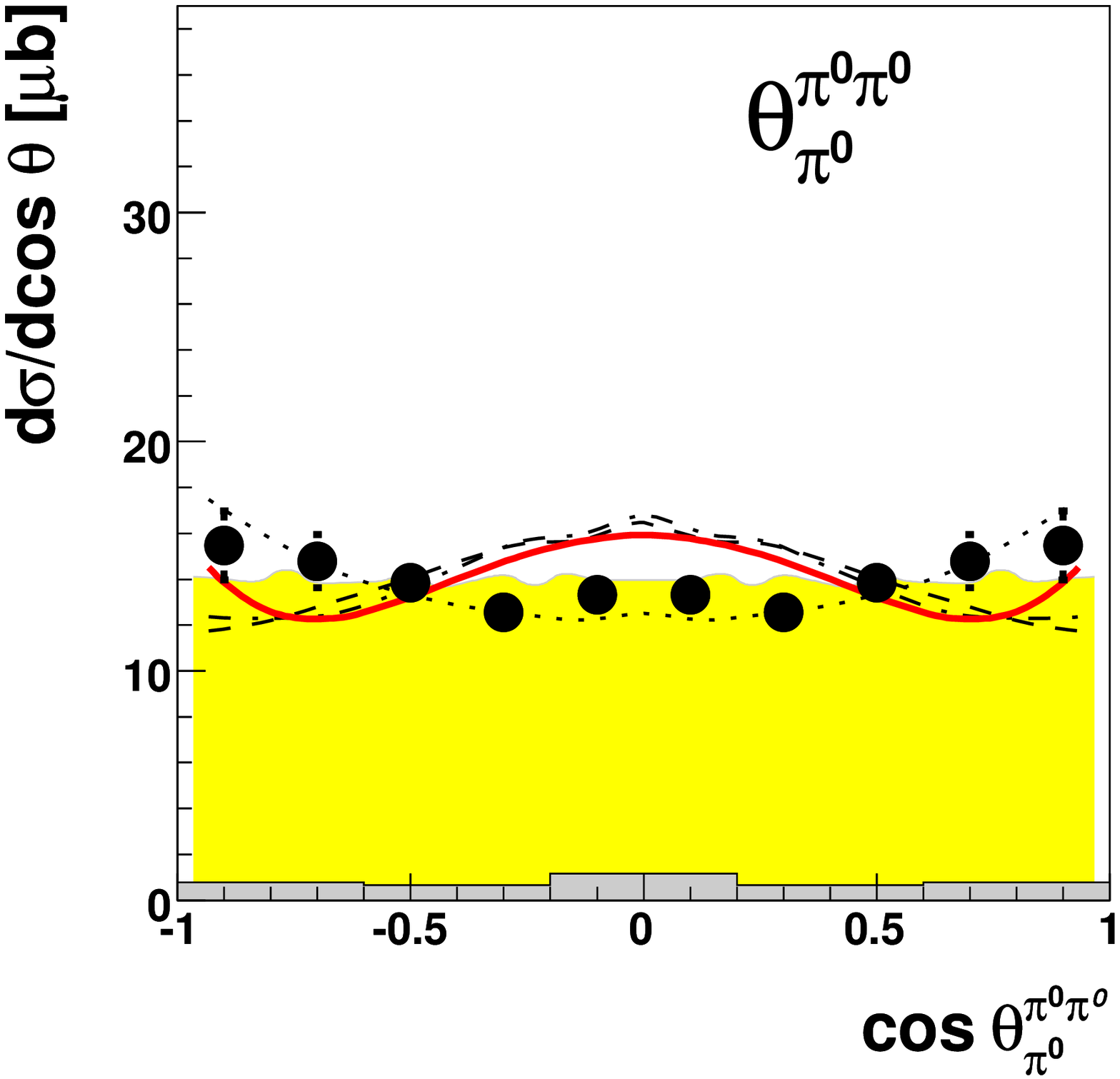}

\includegraphics[width=0.23\textwidth]{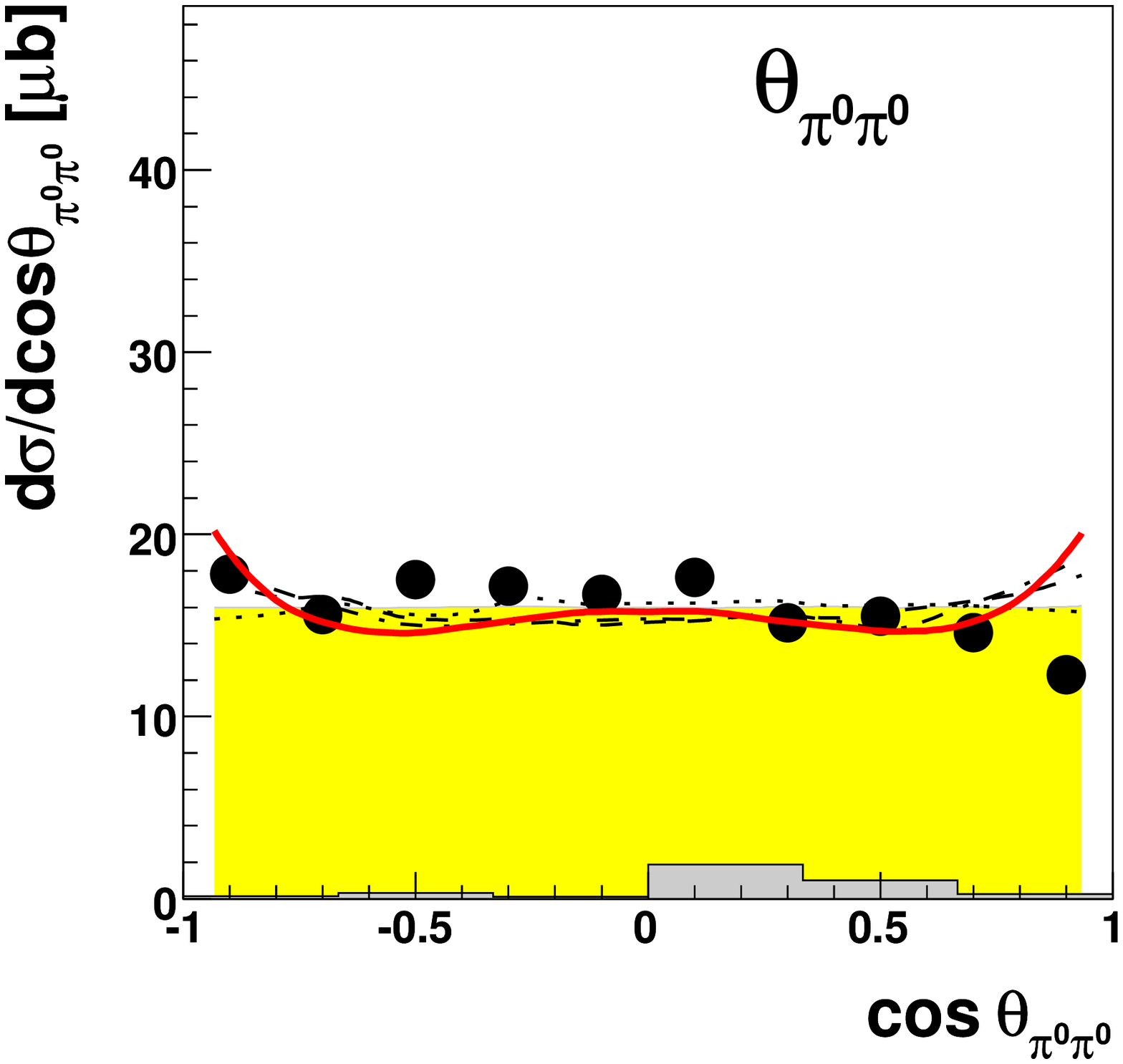}
\includegraphics[width=0.23\textwidth]{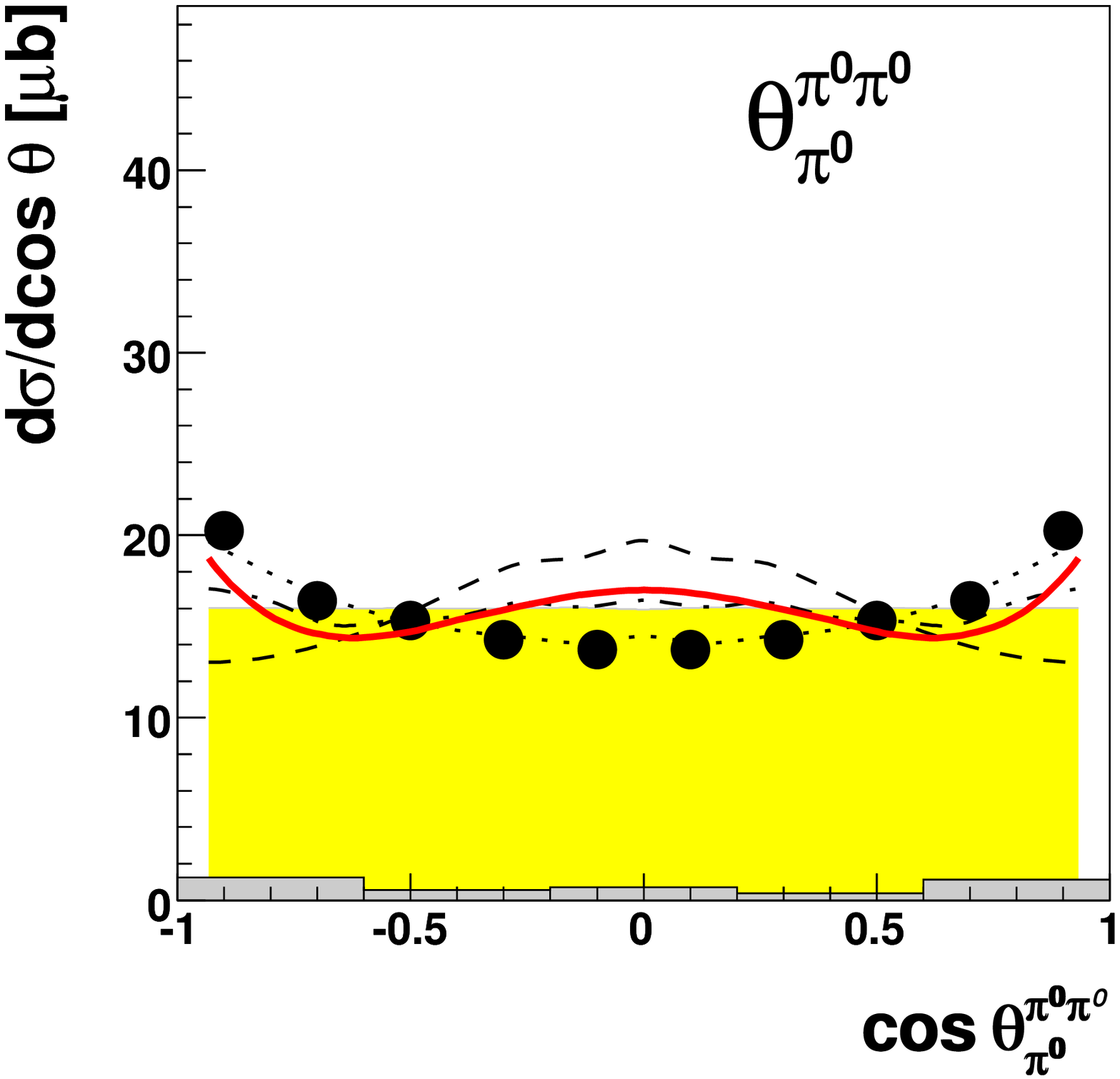}

\includegraphics[width=0.23\textwidth]{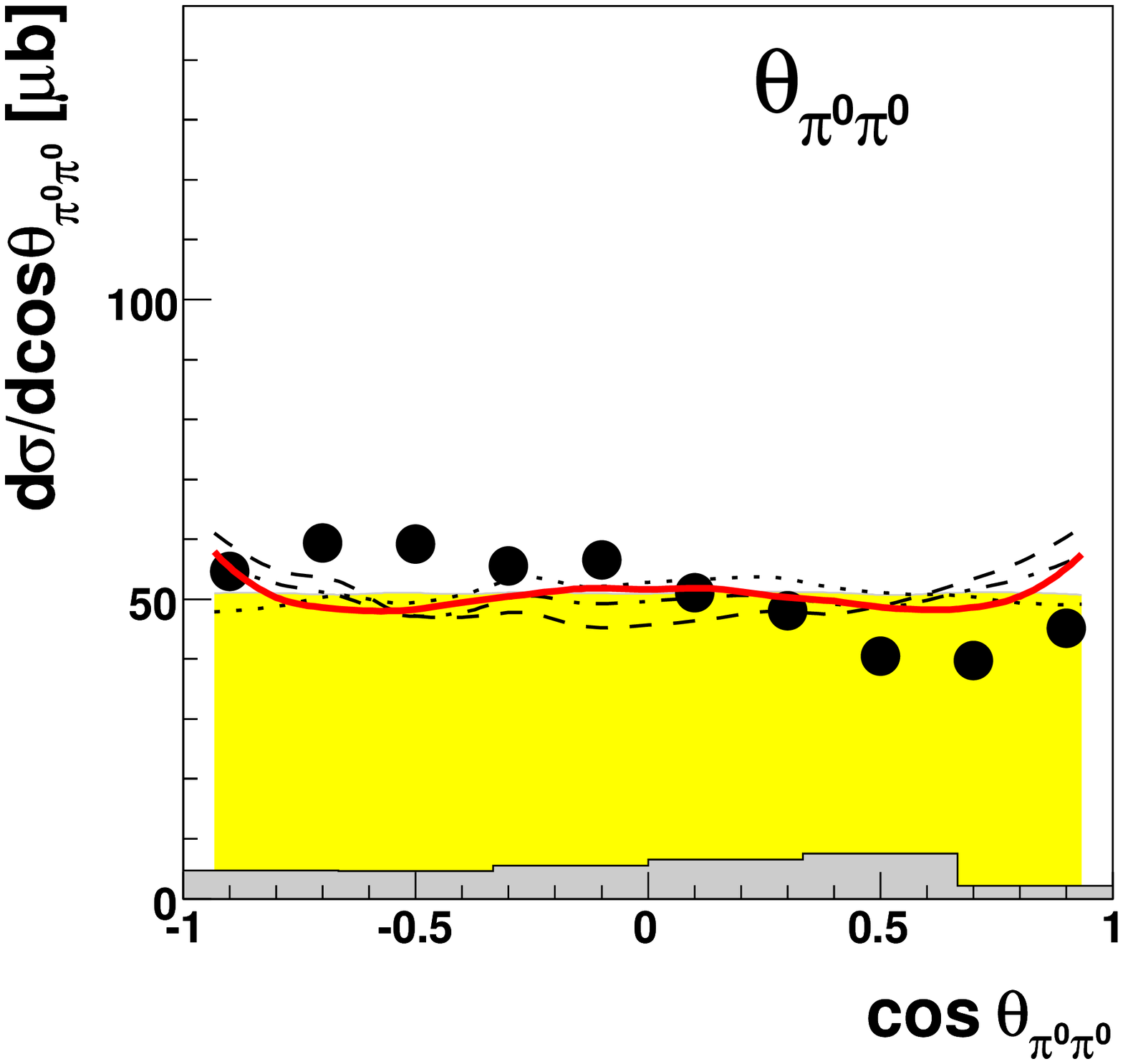}
\includegraphics[width=0.23\textwidth]{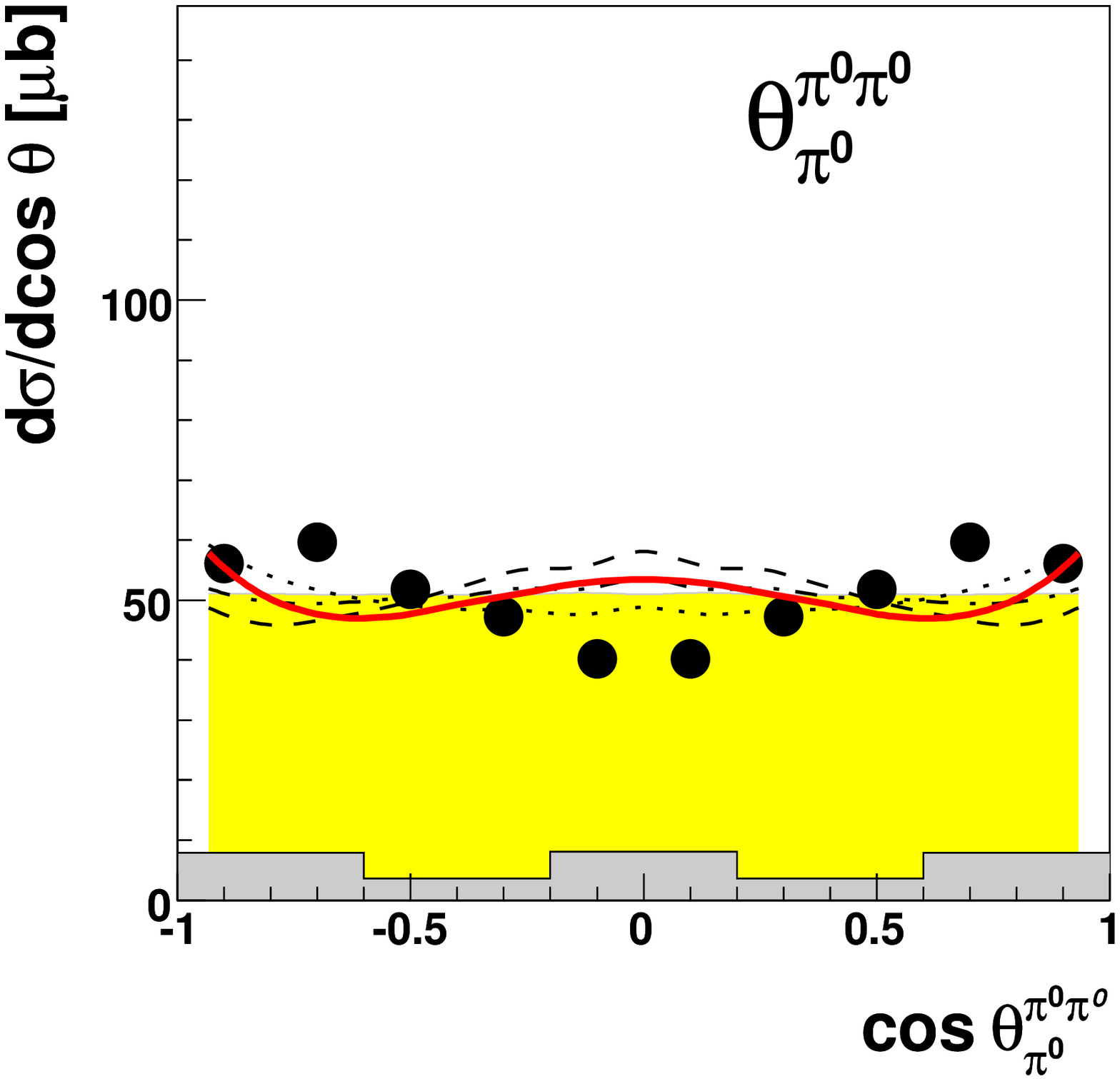}

\caption{
  Same as Fig. 2 but for the 
   distributions of the angle of the $\pi^0\pi^0$ system in the cms
   $\Theta_{\pi^0\pi^0}$  ({\bf left}) and the $\pi^0$ angle in the
   $\pi^0\pi^0$ subsystem $\Theta_{\pi^0}^{\pi^0\pi^0}$  ({\bf right}) for the $pp
   \to pp\pi^0\pi^0$ reaction at beam energies $T_p$ = 1.0, 1.1, 1.2 and
   1.3 GeV (from {\bf top} to {\bf bottom}). 
}
\label{fig4}
\end{center}
\end{figure}

The total cross section data from this work have been published already 
\cite{tsi} in connection with the isospin decomposition of two-pion production
data. They are shown in Fig. 1 together with previous bubble-chamber results
\cite{shim,eis}.  The total cross section keeps rising from
threshold up to $T_p \approx$ 1 GeV, where it levels off and proceeds only
slowly rising  until 1.2 GeV. Thereafter it continues steeply rising until 1.5
GeV, 
where it finally levels off again -- see Fig. 3 in Ref. \cite{tsi}. As
has been demonstrated \cite{tsi}, the low-energy structure is
due to the Roper resonance, whereas the renewed rise at higher energies can be
associated with the dominance of the $\Delta\Delta$ excitation.

In Figs. 2 - 5 we exhibit a selection of eight distributions, which 
are most significant with regard to their physics content. Note that for a
four-body reaction with unpolarized beam and target there are seven 
independent single differential distributions.
For $T_p$ = 1.0, 1.1, 1.2 and 1.3 GeV differential distributions are shown for
the invariant masses $M_{\pi^0\pi^0}$, 
$M_{p\pi^0}$, $M_{pp\pi^0}$ , the opening angle between the two pions
$\delta_{\pi^0\pi^0}$, the proton
angle $\Theta_p$, the $\pi^0$ angle $\Theta_{\pi^0}$, the angle of the
$\pi^0\pi^0$ system $\Theta_{\pi^0\pi^0}$ - all in the cms - as well as the
$\pi^0$ angle in the $\pi^0\pi^0$ subsystem (Jackson frame).  

In Figs. 1 - 5 the data are compared to pure phase space distributions
(light-shaded areas in  Figs. 2 - 7) as well as to calculations of
Ref. \cite{alv} with and without modifications as will be
discussed in the following. As a convention we show all theoretical model 
distributions normalized to the experimental total cross section, {\it i.e.} to
the same area in the differential distributions. This is because we are
interested here in the shape of the differential distributions. The comparison
with the absolute total cross sections is done in Fig. 1.

We see that many of the experimental differential distributions are not far
from the phase space distributions. With regard to angular distributions
significant deviations from isotropy are observed only for the $\Theta_p$ and
$\delta_{\pi^0\pi^0}$ distributions. The first one is largely characterized by
the $t$-channel exchange as demonstrated in Ref. \cite{WB}, whereas the latter
one is strongly correlated with the $M_{\pi^0\pi^0}$ spectrum as discussed in
some detail in Ref. \cite{JP}. In fact, we observe some deviations from phase
space also for the $M_{\pi^0\pi^0}$ spectrum as will be discussed below. However,
really large deviations from phase space are observed in spectra, which are
correlated with $\Delta$ excitation: $M_{p\pi^0}$, $M_{p\pi^0\pi^0}$ and
$M_{pp\pi^0}$. 
The peaks in all these invariant mass spectra build up increasingly with
increasing energy and reflect the increasing excitation of the $\Delta\Delta$
system, 
the resonance pole of which is reached at $T_p  \approx$ 1.3 GeV, i.e. the
highest energy considered in this work. The $M_{pp}$ spectrum, a sample of
which is shown in part in Fig. 6, is kinematically complementary to the
$M_{\pi^0\pi^0}$ spectrum. 

Next we confront the data with model predictions \cite{alv} of the
Valencia group and subsequent modifications of the original calculations. The
dotted lines in Figs. 1 - 5 show the original predictions, which though
renormalized in area give huge deviations from the measured distributions - in
particular in the $M_{\pi^0\pi^0}$ and $\delta_{\pi^0\pi^0}$ spectra. For the
$M_{\pi^0\pi^0}$ spectrum these calculations predict a kind of double-hump
structure with a large enhancement at high invariant masses, which is absent in
the data. With regard to the $\pi^0\pi^0$ opening angle $\delta_{\pi^0\pi^0}$
these calculations predict a preferential antiparallel emission of the two
pions, which again is not supported by the data. 

The predicted double-hump
structure is reminiscent of similar predictions for the ABC effect in the 
double-pionic fusion. There a double-hump structure has been predicted
\cite{ris,anj,colin} based on a $t$-channel $\Delta\Delta$
excitation  --- with some of such calculations \cite{alv2,oster} emphasizing a 
particular strong high-mass enhancement. By contrast exclusive and
kinematically complete measurements of the isoscalar double-pionic fusion find
only a
huge low-mass enhancement, but no or no significant high-mass enhancement
\cite{bash,MB,SK}. As already mentioned above our data for the
$M_{\pi^0\pi^0}$ spectrum are in qualitative agreement with phase space, i.e. we
observe neither a spectacular low-mass enhancement (ABC effect) nor a
spectacular high-mass enhancement. 

In order to shed some light into the failure of the theoretical
predictions, we readjust these calculations step by step first by
implementing the
knowledge accumulated from the studies of the near-threshold region ($T_p <$ 1
GeV), which is governed by excitation and decay of the Roper resonance. From
the analysis of the data at $T_p$ = 0.775 and 0.895 GeV \cite{skor} we obtain
a value for the relative branching between the  
decay via the $\Delta$ resonance, {\it i.e.} $N^* \to \Delta\pi \to
N(\pi\pi)_{I=l=0}$ and the direct decay  $N^* \to  N(\pi\pi)_{I=l=0}$, which
is four times smaller than 
that quoted in PDG \cite{pdg} and used in Ref. \cite{alv}, but in agreement
with a recent analysis of data on pion- and photo-induced pion production on
the nucleon \cite{BoGa}. Updating the model calculations with this new
branching for the Roper decay (dashed lines in Figs. 1 - 5) leads not only to
a quantitative description of the data for $T_p <$ 1 GeV \cite{skor}, but also
to a considerable improvement in the description for $T_p \geq$ 1 GeV, though
there are still substantial deficiencies in the  $M_{\pi^0\pi^0}$ and
$\delta_{\pi^0\pi^0}$ distributions.

Next we readjust the total strength of the Roper excitation.
We know from the isospin decomposition \cite{tsi} of the total two-pion
production cross sections that for $T_p \geq$ 1 GeV the excitation of the Roper
resonance comes out much too strong in the Valencia model calculations: Hence 
we readjust the strength of the Roper excitation
according to the isospin decomposition result. That way we force the
calculations to also reproduce the total cross sections. The outcome of this 
modified calculation is shown in Figs. 1 -
5 by the dash-dotted lines. We now get a good description of the data at $T_p$
= 1.0 GeV. However, deviations from the
data still increase with increasing beam energy. In particular, because the
last modification increased strongly the dominance of $\Delta\Delta$ excitation
for $T_p >$ 1 GeV, we obtain pronounced double-hump and parabolic structures,
respectively, in the  $M_{\pi^0\pi^0}$ and $\delta_{\pi^0\pi^0}$ distributions.

Since these failures are now intimately connected to the treatment of the
$\Delta\Delta$ excitation, further improvements have to be sought in a
modification of this process. As demonstrated already in
Ref. \cite{JP}, double-hump structures in $M_{\pi^0\pi^0}$ spectra are
generated by
a {\bf $\vec{k_1} \cdot \vec{k_2}$} term, where {\bf $\vec{k_1}$}and {\bf
  $\vec{k_2}$} denote the 3-momenta of 
the two emitted pions. In the description of the $\Delta\Delta$ process we
find indeed such a {\bf$\vec{k_1} \cdot \vec{k_2}$} term associated with the
$\rho$ exchange, 
see equation (A.10) in Ref. \cite{alv}. A closer inspection of eq. (A.10) shows
the following problem: the $\rho$ exchange provides isotropic angular
distributions for pions as required by the data, however, provides also a
{\bf$\vec{k_1} \cdot \vec{k_2}$} term at variance with the data. On the
contrary, the pion exchange contains no {\bf$\vec{k_1} \cdot \vec{k_2}$} term,
but involves a strong angular dependence for the pions - in disagreement
with the data. 
Whereas pion-exchange appears to be straightforward
theoretically and well established in the description of pion-production
processes, stringent tests are missing for the $\rho$-exchange. Also since the
$\rho$-exchange part in the $t$-channel $\Delta\Delta$ description of
Ref. \cite{alv} is the by far dominating part, we have to seek for a solution
in the ansatz for the $\rho$ exchange. 

In a recent theoretical work on two-pion production Cao, Zou and
Xu \cite{cao} find that, in contrast to Ref. \cite{alv}, $\rho$-exchange
is small compared to pion-exchange. Triggered by this finding and in view of our
problems to describe the data within the context of the ansatz equation
(A.10) in Ref.\cite{alv}, we investigated the possibility to
achieve a reasonable description of the data by varying the coupling
strength of $\rho$ exchange. 
For simplification, compared to eq. (22) of Ref.\cite{alv}, we use the
longitudinal and transversal exchange contributions to be given purely by
$\pi$- and $\rho$-exchange, respectively. Also for sake of simplicity we drop
the monopole meson form factors. 

In
order to have the description relativistically more appropriate we take the
momenta of the pions from the $\Delta$ decay in the corresponding $\Delta$
systems and correct them by the Blatt-Weisskopf barrier factors as given by
Pilkuhn \cite{pil} and used also in Refs. \cite{ris,cao}. These relativistic
corrections are, of course,  negligible close to threshold, but lead to
sizable and significant changes at higher energies. In particular the
calculated total cross sections no longer keep steeply rising beyond the
resonance pole as in Ref. \cite{alv}, but go into saturation as shown in
Ref. \cite{cao}. The latter behavior agrees with the trend of data at high
energies. 

In fact, we find a quite reasonable description of the data by reducing the
coupling strength of the $\rho$ exchange (A.10) of Ref.\cite{alv} by 
an order of magnitude. The solid lines in Figs. 1 - 5 show our calculations
with the $\rho$ coupling strength reduced by a factor of 12 and reversed in
sign. 
Note that now the $\rho$-exchange is only a small correction
to the leading $\pi$-exchange, but this correction leads still to a sizable
improvement in the description of the differential data. The description of
the data is still far from being perfect, but at least all major features of
the data are reproduced.

Fig. 1 shows the energy dependence of the total cross section of the $pp \to
pp\pi^0\pi^0$ reaction. As mentioned in the introduction the striking
feature is the slow rise of the cross section between $T_p$ = 1.0 and 1.2
GeV, which is reproduced neither by the calculations of Ref. \cite{alv} nor by
the more recent ones of  Ref.\cite{cao}. The primary reason for 
that failure is that in both calculations the Roper excitation keeps rising
beyond $T_p$ = 1.0 GeV, which is at variance with the finding from the isospin
decomposition \cite{tsi}. As pointed out above the differential cross sections
require the Roper excitation to be cut down, too, in accordance with the 
isospin decomposition result. Hence, accounting for the latter and using the
modified $\Delta\Delta$ description we succeed in obtaining
a description for the total cross section, which is in quantitative agreement
with the data.


The comparison to the data shows that the $\rho$
exchange contribution as treated in Ref. \cite{alv,alv2} needs 
substantial modification. This is true both for the two-pion production to
unbound nuclear systems and for the double-pionic fusion. No significant
high-mass enhancement is observed in either case, which would be a signature
of a dominant $\rho$-exchange.  

In the double-pionic fusion experiments a huge low-mass
enhancement (ABC-effect) is found instead, which has been associated with the
formation of an isoscalar resonance via a $\Delta\Delta$ doorway
\cite{MB}. Such an isoscalar resonance can not contribute to the isovector $pp \to pp\pi^0\pi^0$ channel discussed here. The small low-mass enhancement
visible in the data can be fully associated with the conventional $t$-channel
$\Delta\Delta$ excitation. 

\begin{figure} [t]
\begin{center}
\includegraphics[width=0.23\textwidth]{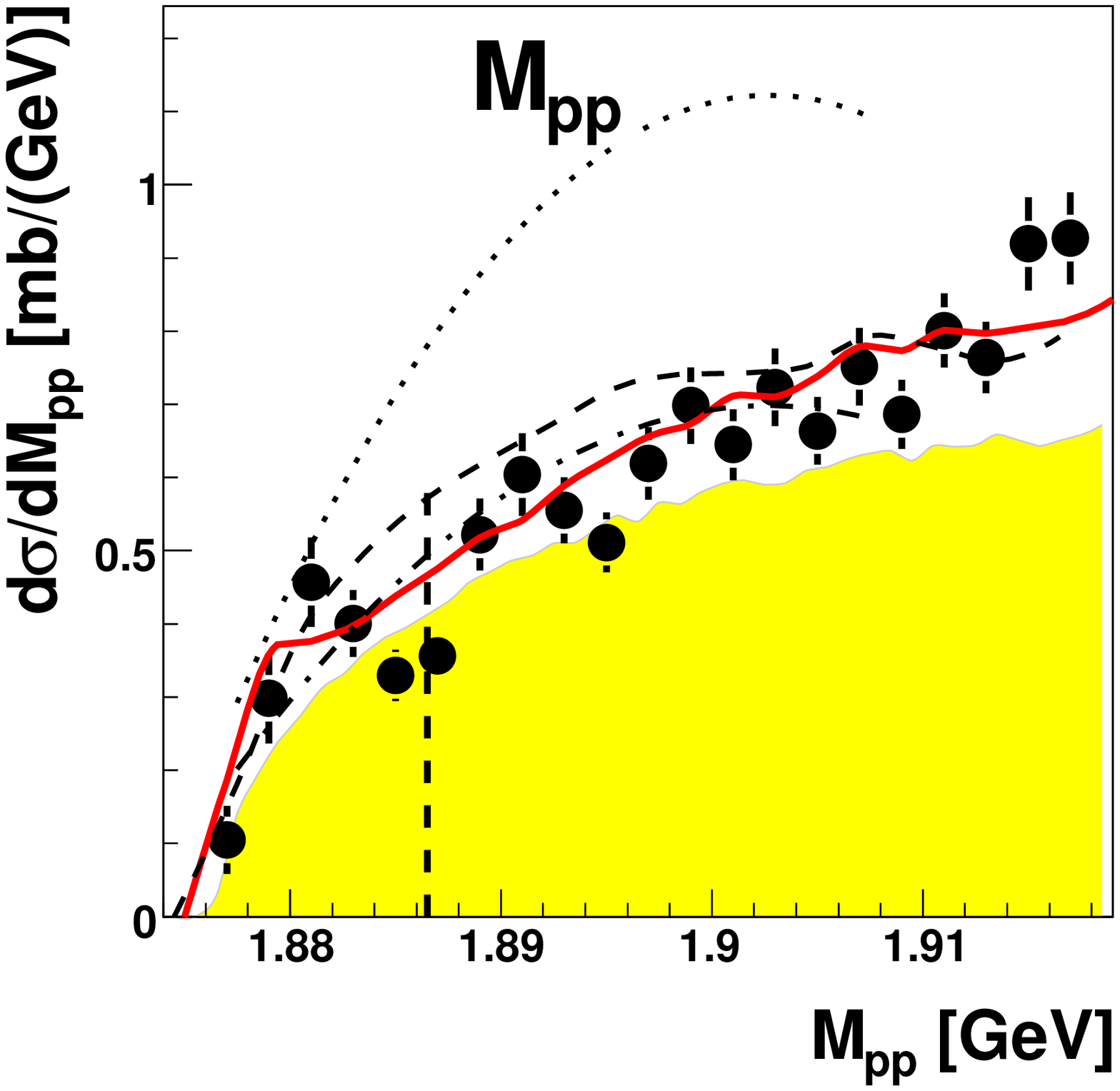}
\includegraphics[width=0.23\textwidth]{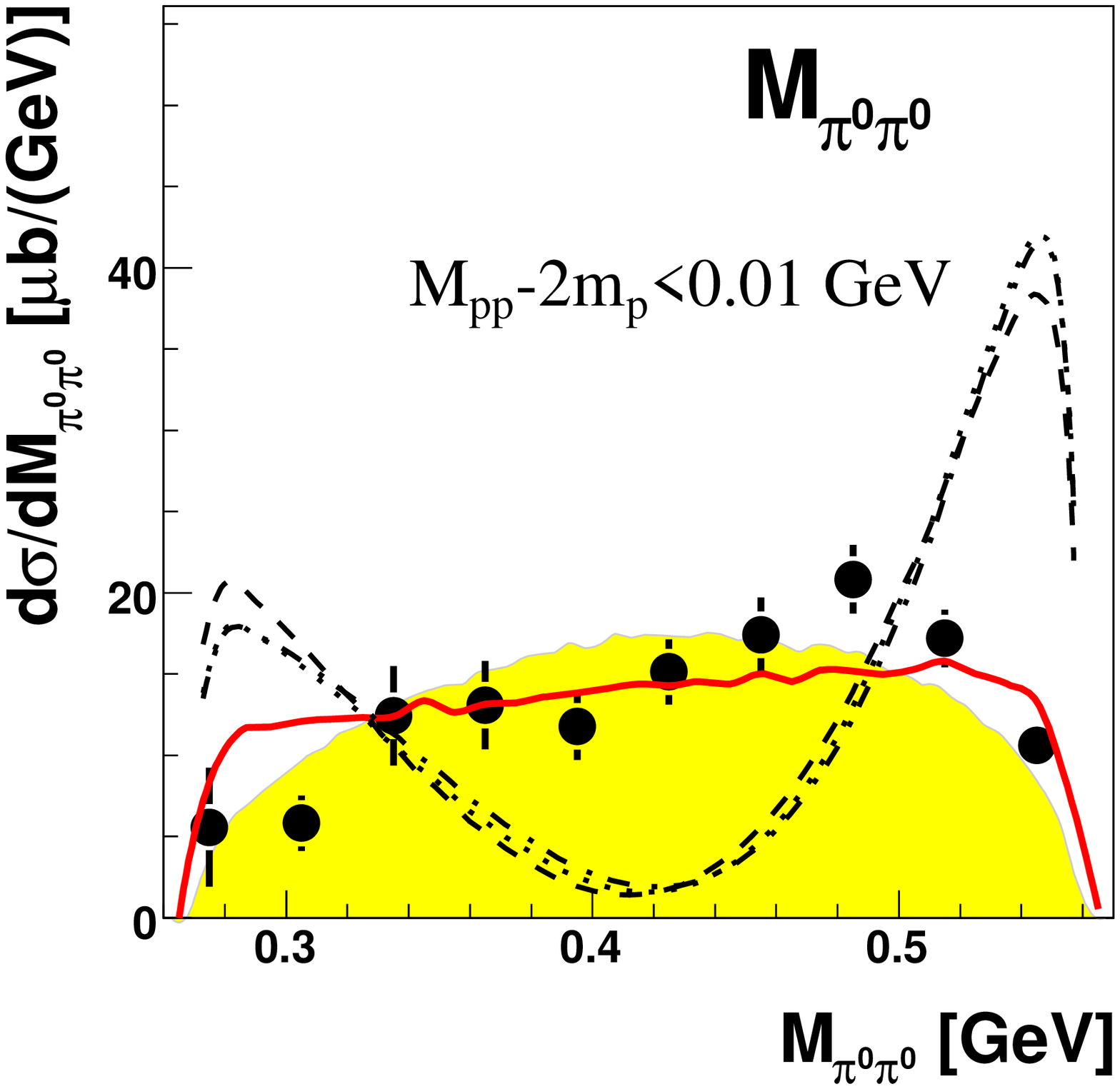}
\includegraphics[width=0.23\textwidth]{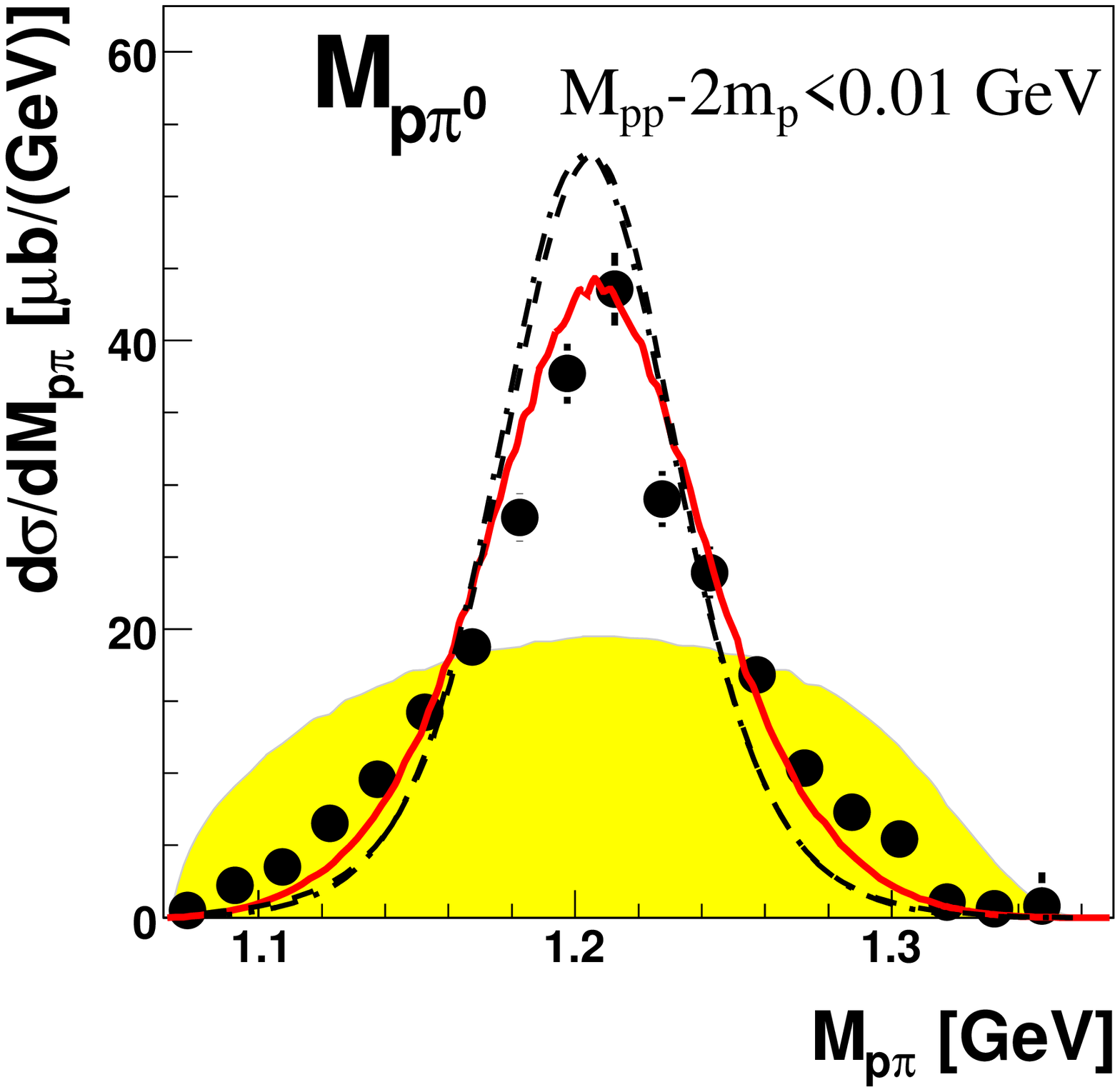}
\includegraphics[width=0.23\textwidth]{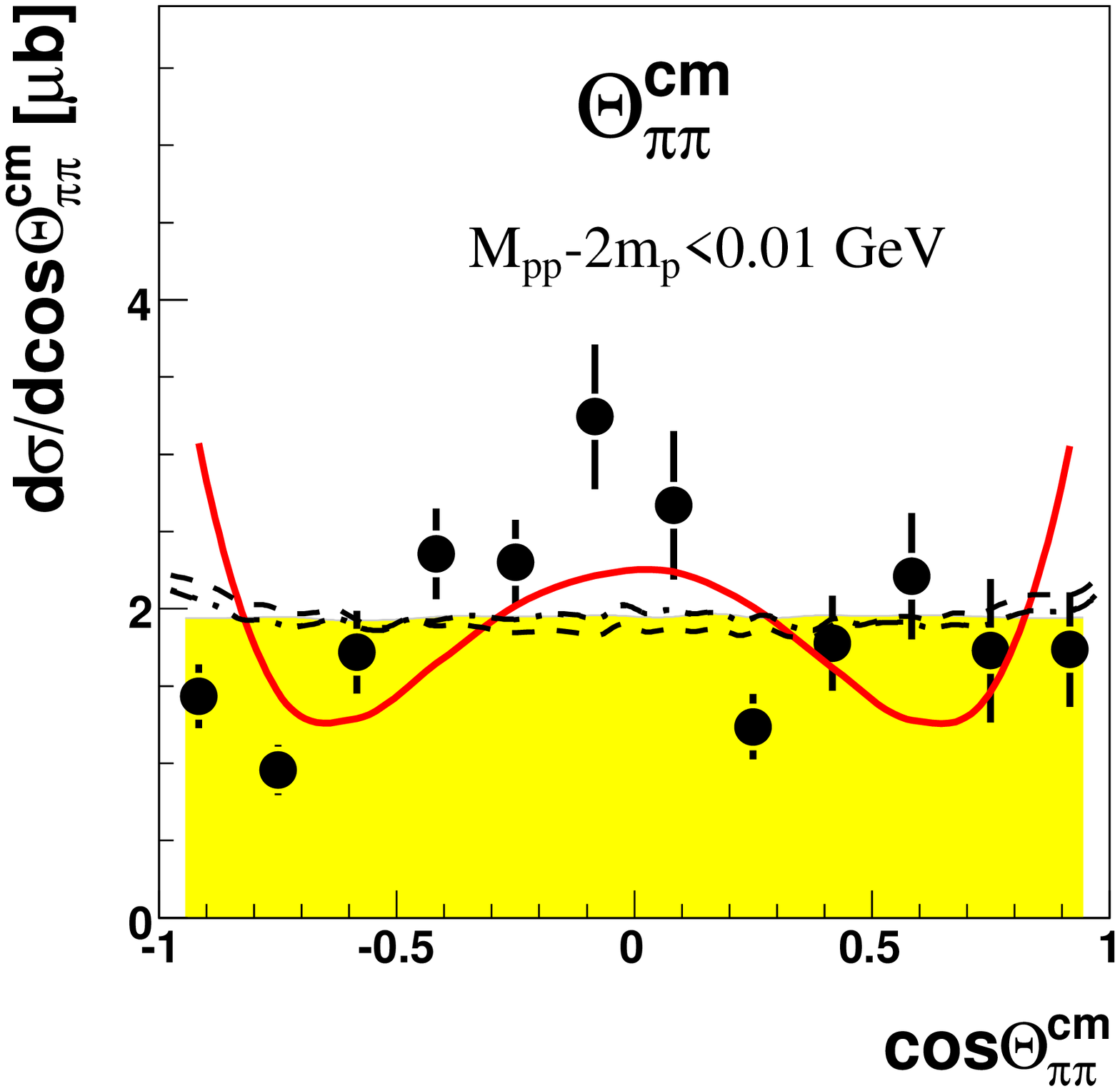}

\caption{
   Differential distributions at $T_p$ = 1.3 GeV for the $^2$He scenario. {\bf
     Top left}: $M_{pp}$ spectrum for the full phase space (only low-mass part
   shown). The distributions
   for   $M_{\pi^0\pi^0}$ ({\bf top right}),  $M_{p\pi^0}$ ({\bf bottom left}) and
 $\Theta_{\pi^0\pi^0}$ ({\bf bottom right}) are plotted with the $^2$He
 condition  $M_{pp} < 2 m_p + 10$ MeV (vertical line in the $M_{pp}$
 spectrum). The  drawn lines represent calculations as described in the caption
 of Fig.\ 2 and text. The shaded areas show the phase space distributions. 
}
\label{fig5}
\end{center}
\end{figure}

In Ref. \cite{dymov} the appealing idea has been put forward that though there
is no bound state in the isovector $NN$ system one could look for a quasibound
$^2$He double-pionic fusion process in the reaction $pp \to pp\pi\pi$ by
requiring that the emitted protons have very small kinetic relative energies. In
Ref. \cite{dymov}, which presents and discusses COSY-ANKE data for the
inclusive reaction $pp \to ppX$, this
requirement was achieved by the condition $M_{pp} \leq 2m_p + 3$ MeV. 
These data --- covering just the forward angle region --- exhibit a pronounced
low-mass enhancement  in the $pp$ missing mass spectrum, which is equivalent
to the associated $\pi\pi$ invariant spectrum.
In our experiment we cover
practically the full phase space of the two-pion production reactions. Though
we have accumulated quite some statistics of events distributed over the full
phase space, we lose nearly all events, if we apply the above $M_{pp}$
cut. Therefore, in order to have acceptable statistics, we relax the above
constraint to $M_{pp} \leq 2m_p + 10$ MeV. In Fig.~6 we show for $T_p$ = 1.3
GeV --- the energy, where the $\Delta\Delta$ process is most obvious in our data
--- the $M_{pp}$ spectrum together with the 
indicated cut as well as the $M_{\pi^0\pi^0}$, $M_{p\pi^0}$ and
 $\Theta_{\pi^0\pi^0}$ spectra resulting from this cut. In the $M_{pp}$
 spectrum we observe an indication of the low-mass enhancement produced by the
 $pp$ final-state interaction, which is also included in the theoretical
 calculations.  In order to focus on small masses we plot this spectrum only
 in the low-mass range in Fig. 6.  We see that the cut $M_{pp} \leq 2m_p + 10$
 MeV is still within the region dominated by the $pp$
 final-state interaction ensuring thus relative s-waves between the two
 protons. 
Within the limited statistics the data in the constrained
$M_{\pi^0\pi^0}$ spectrum  exhibit some high-mass enhancement as 
expected from the pioneering work of Risser and Shuster \cite{ris} concerning
the production of an isoscalar pion pair via the $\Delta\Delta$ excitation in
the double-pionic fusion. The constrained
$M_{p\pi^0}$ spectrum exhibits the $\Delta$ peak ensuring the
$\Delta\Delta$ process to be the dominating process in the  $^2$He case, too.

The solid
lines, which present the modified theoretical description of the $pp \to
pp\pi^0\pi^0$ reaction, give a reasonable description for this constrained phase
space scenario. On the contrary, the broken lines exhibit a very pronounced
double-hump structure in the $M_{\pi^0\pi^0}$ spectrum due to the dominance of
the $\rho$-exchange in these calculations. This behavior is clearly at
variance with the data. On basis of the model calculation we also ensure that
the change in 
the cut from a 3 MeV to a 10 MeV range does not change the results
qualitatively. The $\Theta_{\pi\pi}^{cm}$ angular distribution points to some
sizable d-wave contribution, which is in support of the respective
ansatz in Ref. \cite{dymov}. The increased d-wave contribution as compared to
the results for the full $pp \to pp\pi^0\pi^0$ reaction is not unexpected,
since the $^2$He cut forces the two emerging protons to be in relative
s-wave. The decay of the two excited $\Delta$ states results in a double
p-wave emission of the two pions. This in turn leads to s- and d-waves in the
emitted $pp\pi^0\pi^0$ system.
 
Finally we compare our calculations to the COSY-ANKE data in Fig. 7 by 
restricting the calculations further to the angular range covered by the ANKE
experiment, which is $cos \Theta_{pp}  > 0.95$ with $\Theta_{pp} = 180^{\circ}
- \Theta_{\pi^0\pi^0}$ and $\Theta_p^{lab} \leq 12^{\circ}$. We again obtain
an essentially quantitative agreement with the data at $T_p$ = 1.1 and
1.4 GeV using the modified model description, whereas the $\rho$-exchange
dominated calculations again give a double-hump structure in vast
disagreement with the data.  


\begin{figure} 
\begin{center}

\includegraphics[width=0.23\textwidth]{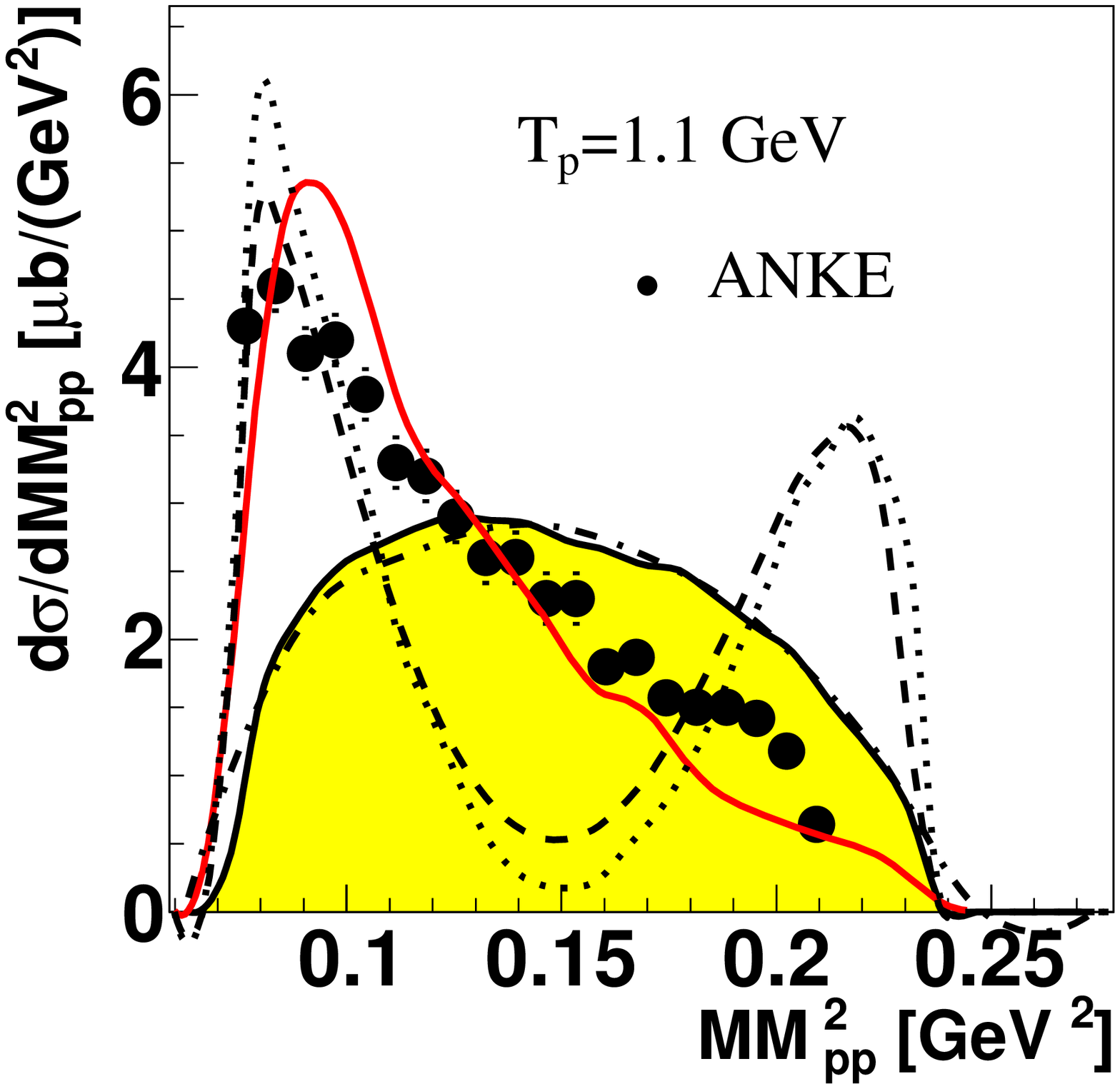}
\includegraphics[width=0.23\textwidth]{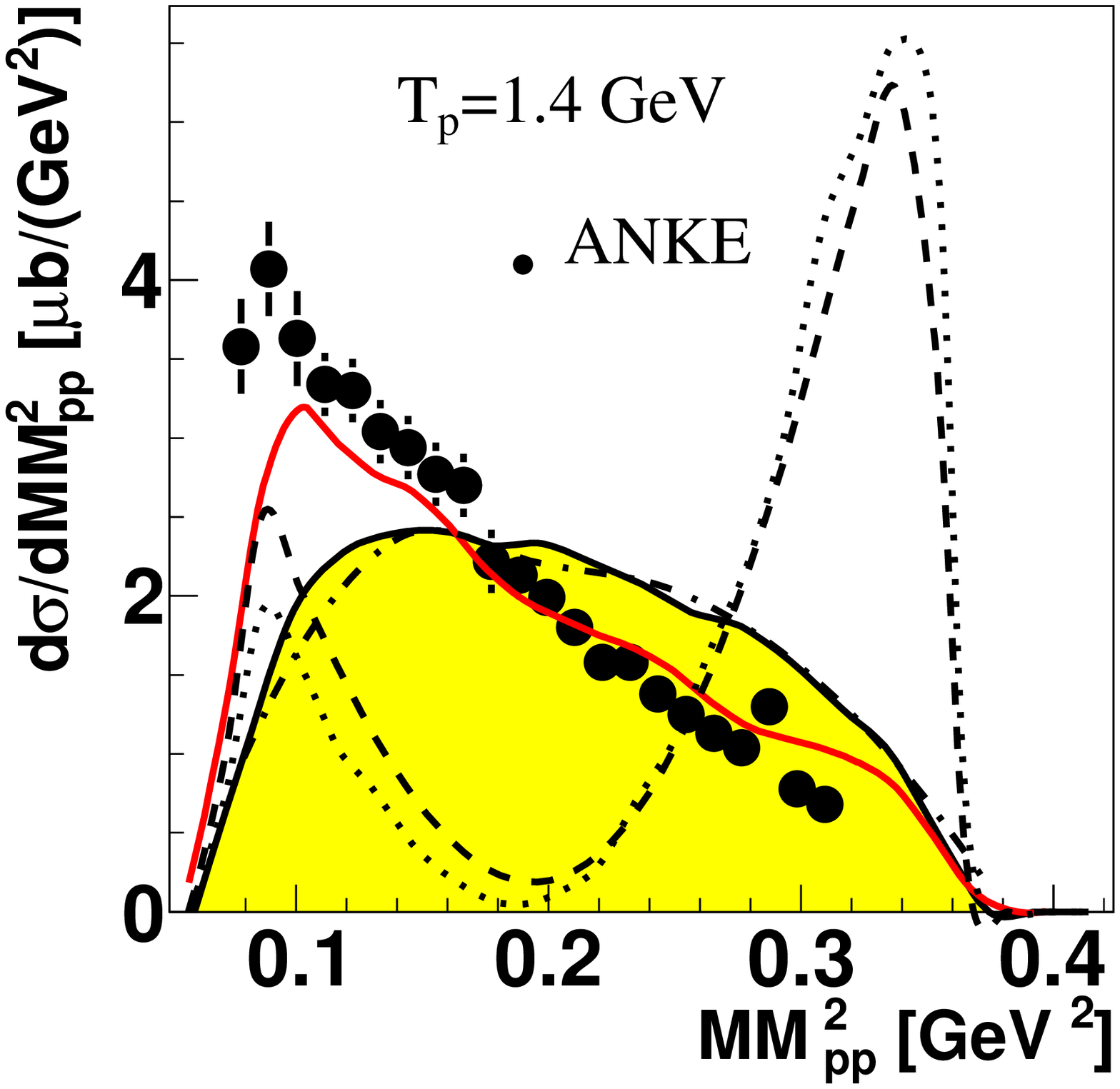}

\caption{
   $M_{\pi^0\pi^0}$ spectra  at $T_p$ = 1.1 and 1.4 GeV for the  $^2$He
 condition  $M_{pp} < 2 m_p + 3$ MeV and the ANKE angular constraints $cos
 \Theta_{pp}  > 0.95$ with $\Theta_{pp} = 180^{\circ} - \Theta_{\pi^0\pi^0}$
 and $\Theta_p^{lab} \leq 12^{\circ}$. The solid dots represent the ANKE data
 \cite{dymov} and the drawn lines calculations as described in the caption of
 Fig.\ 2 and text. The shaded areas show the phase space distributions.
}
\label{fig6}
\end{center}
\end{figure}


Summarizing, we have presented first exclusive and kinematically complete
measurements of the $\Delta\Delta$ system excited in the $pp \to
pp\pi^0\pi^0$ reaction. The data are well described by a conventional
$t$-channel  
$\Delta\Delta$ calculation, where the $\rho$ exchange contribution is strongly
reduced compared to that in Ref. \cite{alv}. The same holds for the
subset of data, which corresponds to the double-pionic fusion to a quasibound
$^2$He nucleus. No evidence is found for a significant low-mass enhancement in
the $M_{pi^0\pi^0}$ spectra (ABC effect) 
beyond that given by the conventional $t$-channel $\Delta\Delta$
excitation process. 
The new data sets should serve as a significant test case for
theoretical treatments of multipion production. 


We acknowledge valuable discussions with L. Alvarez-Ruso, C. Hanhart, E. Oset,
A. Sibirtsev and C. Wilkin on this issue. We are particularly indebted to
L. Alvarez-Ruso for using his code.  
This work has been supported by BMBF
(06TU9193), Forschungszentrum J\"ulich (COSY-FFE) and  
DFG (Europ. Graduiertenkolleg 683). 

\end{document}